\documentclass[fleqn,10pt]{wlscirep}
\usepackage[utf8]{inputenc}
\usepackage[T1]{fontenc}
\usepackage{bm}
\usepackage{graphicx}
\usepackage{amsmath}
\usepackage{amsfonts}
\usepackage{amssymb}
\usepackage{caption}
\usepackage{subcaption}
\usepackage{todonotes}
\usepackage{biblatex}
\usepackage{siunitx}
\usepackage{xspace} 
\usepackage{authblk}
\usepackage{multirow}
\usepackage{csquotes}
\long\def\/*#1*/{}
\newcommand{\AI}[1][]{AI algorithm#1\xspace}
\newcommand{\network}{deep regression network\xspace}
\newcommand{\networkembedding}{feature embedding network\xspace}
\newcommand{\celldetector}{epithelial cell detection network\xspace}
\newcommand{\ROIdata}{ROI data set\xspace}

\newcommand{\mMS}[1]{\ensuremath{\mathbb{#1}}} % multiple slides/slide space
\newcommand{\mSS}[1]{\ensuremath{\mathcal{#1}}} % single slide/slide space
\newcommand{\mMR}[1]{\ensuremath{\mathcal{#1}}} % multiple regions/region space
\newcommand{\mMI}[1]{\ensuremath{\mathbf{#1}}} % multiple items(patches)/item(patch) space
\title{Automated Scoring of Nuclear Pleomorphism Spectrum with Pathologist-level Performance in Breast Cancer}

\author[a,*]{Caner Mercan}
\author[a]{Maschenka Balkenhol}
\author[b,c]{Roberto Salgado}
\author[d]{Mark Sherman}
\author[e]{Philippe Vielh}
\author[f]{Willem Vreuls}
\author[g,]{Ant\'{o}nio Pol\'{o}nia}
\author[h]{Hugo M. Horlings}
\author[i]{Wilko Weichert}
\author[d]{Jodi M. Carter}
\author[a]{Peter Bult}
\author[j]{Matthias Christgen}
\author[k]{Carsten Denkert}
\author[l]{Koen van de Vijver}
\author[a,m,$\dag$]{Jeroen van der Laak}
\author[a,$\dag$]{Francesco Ciompi}
\affil[a]{%
    Radboud University Medical Center, 
    Radboud Institute for Health Sciences, 
    Department of Pathology, 
    Nijmegen, The Netherlands}
\affil[b]{GZA-ZNA Hospitals, Department of Pathology, Antwerp, Belgium} % Roberto
\affil[c]{Peter Mac Callum Cancer Centre, Division of Research, Melbourne, Australia} % Roberto
\affil[d]{Mayo Clinic, Department of Laboratory Medicine and Pathology, Rochester, Minnesota, USA} % Mark, Jodi
\affil[e]{Medipath \& American Hospital of Paris, Paris, France} % Philippe
\affil[f]{Canisius Wilhelmina Ziekenhuis, Nijmegen, The Netherlands} % Willem
\affil[g]{%
    University of Porto,
    Institute of Molecular Pathology and Immunology
    Department of Pathology, 
    Ipatimup Diagnostics, 
    Porto, Portugal
}
\affil[h]{%
    The Netherlands Cancer Institute,
    Department of Molecular Pathology,
    Amsterdam, The Netherlands} %Hugo
\affil[i]{Technical University Munich, Institute of Pathology, Munich, Germany} %Wilko
\affil[j]{Hannover Medical School, Institute of Pathology, Hannover, Germany} %Matthias 
\affil[k]{Philipps University of Marburg, Institute of Pathology, Marburg, Germany} %Carsten
\affil[l]{%
    Ghent University Hospital and Cancer Research Institute Ghent,
    Department of Pathology, 
    Ghent, Belgium} %Koen
\affil[m]{Center for Medical Image Science and Visualization, Linköping University, Linköping, Sweden}
\affil[$\dag$]{shared senior authorship}
\affil[*]{e-mail: caner.mercan@radboudumc.nl}

\begin{abstract}
Nuclear pleomorphism, defined herein as the extent of abnormalities in the overall appearance of tumor nuclei, is one of the components of the three-tiered breast cancer grading, along with degree of gland formation and mitotic count.
The degree of nuclear pleomorphism is subjectively classified from 1-3, where a score of 1 most closely resembles epithelial cells of normal breast epithelium and 3 shows the greatest abnormalities.
In contrast to gland formation and mitotic count, which are evaluated according to quantitative criteria, establishing numerical criteria for grading nuclear pleomorphism is challenging, and inter-observer agreement is poor.
Given that nuclear pleomorphism reflects a continuous spectrum of variation, we trained a deep neural network on a large variety of tumor regions from the collective knowledge of several pathologists, without constraining the network to the traditional three-category classification.
We also motivate an additional approach in which we discuss the additional benefit of normal epithelium as baseline, following the routine clinical practice where pathologists are trained to score nuclear pleomorphism in tumor, having the normal breast epithelium for comparison.
In multiple experiments, our fully-automated approach could achieve top pathologist-level performance in select regions of interest as well as at whole slide images, compared to ten and four pathologists, respectively.
\end{abstract}

\begin{document}
% \flushbottom
\maketitle

\section{Introduction}
\label{section:introduction}

A projected 276,480 women will be diagnosed with breast cancer in the U.S. in 2020, and 42,170 women will die of the disease \cite{Sieg20}.
To guide management, most pathologists grade breast cancers according to a standardized grading system \cite{Bloo51, Elst91}, comprised of three features: 1) nuclear pleomorphism (or “atypia”), 2) extent of gland formation and 3) mitotic count.
The scoring criteria for mitotic count and gland formation are defined by quantitative measures, whereas nuclear pleomorphism scoring is based on qualitative analysis of the nuclear morphology of tumor as assessed microscopically on a scale of 1 to 3, reflecting increasing differences in appearance compared with normal epithelium.
The final tumor grade is derived from these three scores.
With the increased utilization of digital pathology, grading can now be carried out on digitized histopathological images, which yields similar agreement as compared to assessment via light microscopy \cite{Gint20}.
Increasingly, the pathologists can also be assisted by Artificial Intelligence (AI) based systems in routine practice \cite{Bera19, Niaz19, Bult20a, Rant20}.

The most commonly applied AI technology for analysis of medical images are so-called deep neural networks (deep learning; DL).
Deep learning architectures are composed of connected neurons that receive an input image and perform a series of operations on the learnable network weights to accomplish a learning task, such as classification, regression and segmentation.
They have led to unprecedented success in several areas of computer vision \cite{Kriz12, Le13, Silv16, Devl18} to more recent advancements in digital pathology \cite{Jano16, Bejn17, Veta18, Nagp19, Bult20}. 
More specific for breast cancer grading, automated mitosis detection was one of the first applications of DL \cite{Cire13a, Veta15}, demonstrating its clinical and prognostic value \cite{Tell18, Balk19, Balk19b}.
Gland segmentation was formulated as a series of nuclei and gland detection, as well as a segmentation task \cite{Nguy15, Lee18}, with extensions to clinical risk categories \cite{Buch16} through the use of deep learning.
Unlike gland formation and mitotic count, nuclear pleomorphism does not have the same quantitative nature in its definition.
As a result, nuclear pleomorphism scoring is the least reproducible of the three grading components, which limits its utility \cite{Frie95, Meye05, Long05}.
The works addressing automated nuclear pleomorphism scoring have been limited, consisting of techniques involving manually crafted nuclear features \cite{Lu15, Gand19} combined with the use of deep belief networks \cite{Maql16}, descriptor-based pixel-level image features \cite{Khan15} or binary classification of high pleomorphism (score 3) using a simple convolutional neural network on small image patches \cite{Noel15}.

Supervised DL requires ground-truth labels associated with the images for neural network training.
Acquiring such labels is a non-trivial task for applications such as pleomorphism scoring, with discrete classes and observer variability.
Translating the scores of multiple observers into a ground-truth label (e.g. based on majority, or consensus) disregards valuable information which is present in the spread of the scores of individual experts.
In this work, we therefore formulate a novel DL approach to nuclear pleomorphism scoring by considering it as a continuous score, rather than classifying it into three discrete categories.
We re-formulate the original three-category classification into a full spectrum of continuous values from the lowest pleomorphism score to the most severe.
Our approach mainly consists of two parts.
The first part is an epithelial cell detection network developed previously \cite{Merc19}.
This step is intended to limit the analysis to the diagnostically relevant regions (i.e. invasive tumor) within a whole slide image.
In the second step, a deep regression network predicts the continuous nuclear pleomorphism score on the tumor regions.
Training this network does not require detailed manual annotations by pathologists which is one of the key limiting factors within computational pathology.
This work marks the first end-to-end fully automated nuclear pleomorphism scoring in breast cancer using DL.
Moreover, we are translating a discrete classification into a continuous regression problem to preserve valuable observer input which is not applied previously.

\begin{figure}
\centering
    \begin{minipage}{0.25\textwidth}
    \centering
        \frame{\includegraphics[width=\columnwidth]{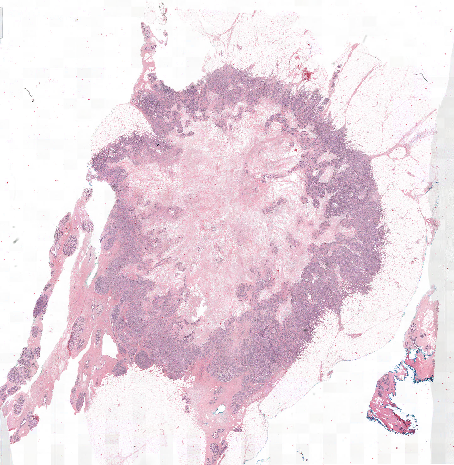}}
        \subcaption{Input slide}
    \end{minipage}%%%
    \begin{minipage}{0.10\textwidth}
    \centering
        \includegraphics[width=\columnwidth]{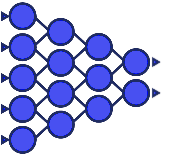}
        % (b) 
    \end{minipage}%%%    
    \hspace{-5pt}
    \begin{minipage}{0.25\textwidth}
    \centering
        \frame{\includegraphics[width=\columnwidth]{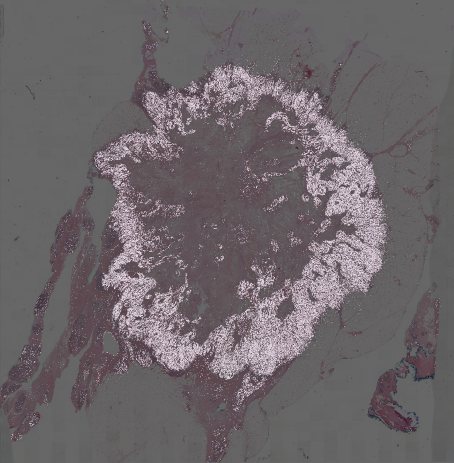}}
        \subcaption{Tumor output}
    \end{minipage}%%%
    \begin{minipage}{0.10\textwidth}
    \centering
        \includegraphics[width=\columnwidth]{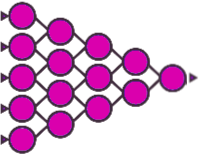}
        % (d)
    \end{minipage}%%%
    \begin{minipage}{0.25\textwidth}
    \centering
        \frame{\includegraphics[width=\columnwidth]{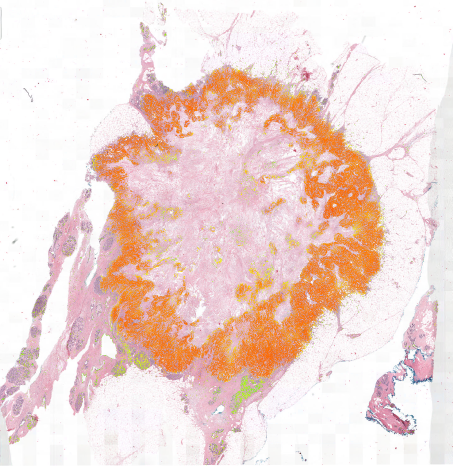}}
        \subcaption{Pleomorphism spectrum}
    \end{minipage}
    \hspace{-4pt}%%%
    \begin{minipage}{0.03\textwidth}
    \centering
        \frame{\includegraphics[height=100pt, width=5pt, trim=15 10 40 10, clip]{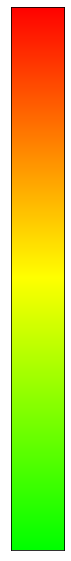}} \\
        \vspace{13pt}
    \end{minipage}
\caption{The overview of the \AI scoring nuclear pleomorphism spectrum. An input whole slide image (a) is processed through the \celldetector to detect the tumor regions. Subsequently, the \network scores nuclear pleomorphism on the tumor regions, score $1$ is denoted by green, score $2$ by yellow and score $3$ by red color. Since the \network outputs a spectrum of values, ranging from $1$ to $3$, reflecting the scores in between the traditional three categories, as well. In this example, the \AI scored high pleomorphism for the tumor regions in the input slide, evident by the dark orange and red colors in the pleomorphism spectrum.}
\label{fig:introduction:overview_inference}
\end{figure}

The overview of our approach, \AI, outlining this two-stage process of nuclear pleomorphism scoring is presented in Figure \ref{fig:introduction:overview_inference}.
In order to develop and validate our approach, we carried out two separate studies, resulting in two separate data sets.
The first study, referred to as ROI-study, consisted of $125$ tumor regions of interest (tumor ROIs) and an additional $79$ "normal appearing" epithelial regions of interest (normal ROIs) from $39$ whole slide images, carefully selected to include a wide variety of tumor morphology in breast cancer histopathology.
We invited $10$ pathologists from $6$ countries to score nuclear pleomorphism on the tumor ROIs, each visually paired with a normal ROI from the same patient for the pathologists to use as reference for scoring the degree of pleomorphism.
We used the results of the ROI-study to train our \AI as well as to evaluate its performance compared to the panel of pathologists.
In our automated approach, we used the averaged nuclear pleomorphism scores over the 10 pathologists as reference scores, representing their collective knowledge rather than forcing a discrete majority score.
The \AI was trained on a wide range of tumor morphology, leveraging the reference scores to establish the concept of nuclear pleomorphism as a full spectrum of tumor morphology change.
In the second study, referred to as the slide-study, $118$ whole slide images of breast cancer resection tissue sections were used for the evaluation of the trained \AI.
The distribution of the slides resembled the distribution of the cases in routine clinical practice.
For the slide-study, we invited $4$ pathologists to score nuclear pleomorphism of the slides which were compared to our \AI for whole slide-level evaluation.
More details on the ROI-study and slide-study are provided in Section \ref{sec:methodology}.

\section{Results}
\label{section:results}
\paragraph{Is tumor morphology enough for automated nuclear pleomorphism scoring?}
In routine clinical practice, pathologists typically score nuclear pleomorphism of tumor via a visual comparison to a region with normal cells as a reference.
In this section, we investigate whether the tumor morphology by itself is sufficient for the \AI to learn the full pleomorphism spectrum without having seen examples from healthy epithelium.
We compared two approaches: 1) learning only from tumor morphology (a \AI trained and validated only on tumor patches from the training and validation sets in the ROI-study), and 2) incorporating ROIs containing normal epithelium on top of the first approach (see Section \ref{sec:methodology} for details).
% The former was an \AI trained and validated only on tumor patches from the training and validation sets in the ROI-study.
% For the second approach, we also incorporated the normal ROIs to sample patches with normal cells (see Section \ref{sec:methodology} for details).
We compared the best models from both approaches on four independent sets of $1000$ patches randomly sampled from the validation set.
For each input patch, the predictions from both models were nuclear pleomorphism scores of continuous values between $[1-\epsilon, 3+\epsilon]$, with $\epsilon$ denoting small prediction deviations outside of the score spectrum.
% We evaluated both models on several regression metrics (mean absolute error ($MAE$), mean squared error ($MSE$), median absolute error ($MdAE$), explained variance score ($EV$) and $R^2$ score) in comparison to the reference scores, as presented in Table \ref{tab:results:normal:network}.
We evaluated both models on several regression metrics (mean absolute error ($MAE$), mean squared error ($MSE$) and explained variance score ($EV$)) in comparison to the reference scores, as presented in Table \ref{tab:results:normal:network}.
The results highlighted that the \AI did not benefit from being exposed to normal epithelium, and clearly learned the nuclear pleomorphism score from the tumor morphology alone (see Figure \ref{fig:data:roi:grade_distribution}).
In the rest of the paper, we report the results from the \AI that was trained only on tumor morphology.
\begin{table}[]
    \centering
    \begin{tabular}{lccccc}
    %  & $MAE \downarrow$ & $MSE \downarrow$ & $MdAE \downarrow$ & $EV \uparrow$ & $R^2 \uparrow$ \\
    % \hline \hline
    % \AI & $ 0.262 \pm 0.004 $ & $ 0.111 \pm 0.002 $ & $ 0.215 \pm 0.008 $ & $ 0.758 \pm 0.008 $ & $ 0.756 \pm 0.009 $ \\
    % \AI with baseline & $ 0.275 \pm 0.006 $ & $ 0.116 \pm 0.007 $ & $ 0.237 \pm 0.014 $ & $ 0.744 \pm 0.014 $ & $ 0.743 \pm 0.016 $ \\
    % \hline 
    % \end{tabular}
    % \caption{Evaluation of the added value of exposure to normal epithelium for the \AIs as a baseline, on a subset of the patches in the validation set of the ROI-study. \AI refers to learning from tumor morphology whereas \AI with baseline refers to the extension in which normal epithelium is used as baseline. Performance is better when mean absolute error ($MAE$), mean squared error ($MSE$) and median absolute error ($MdAE$) are low, explained variance score ($EV$) and $R^2$ score are high. }
    & $MAE \downarrow$ & $MSE \downarrow$ & $EV \uparrow$ \\
    \hline \hline
    \AI & $ 0.262 \pm 0.004 $ & $ 0.111 \pm 0.002 $ & $ 0.756 \pm 0.009 $ \\
    \AI with baseline & $ 0.275 \pm 0.006 $ & $ 0.116 \pm 0.007 $ & $ 0.743 \pm 0.016 $ \\
    \hline 
    \end{tabular}
    \caption{Evaluation of the added value of exposure to normal epithelium for the \AI[s] as a baseline, on a subset of the patches in the validation set of the ROI-study. \AI refers to learning from tumor morphology whereas \AI with baseline refers to the extension in which normal epithelium is used as baseline. Performance is better when mean absolute error ($MAE$) and mean squared error ($MSE$) are low, explained variance score ($EV$) is high.}
    \label{tab:results:normal:network}
\end{table}

\begin{figure}
\centering
    \begin{minipage}{0.33\textwidth}
        \centering
        \frame{\includegraphics[width=1\columnwidth]{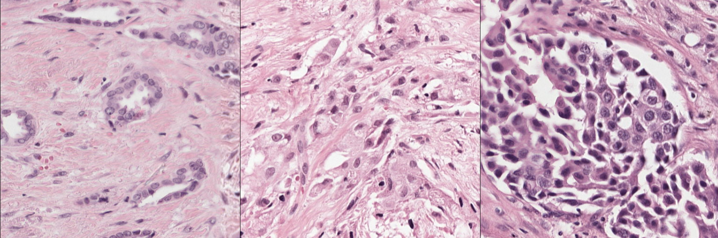}} \\
        \vspace{3pt}
        \frame{\includegraphics[width=1\columnwidth]{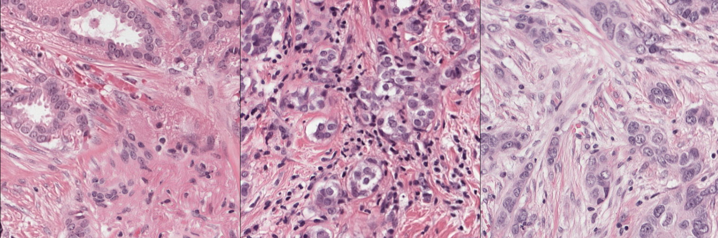}}
        \frame{\includegraphics[width=5pt, height=\columnwidth, angle=270, trim=40 15 10 10, clip]{figures/results/pleo_colorbar.png}} \\
        \vspace{3pt}
        (a)
    \end{minipage}%%%
    \hspace{0.10\textwidth}
    \begin{minipage}{0.55\textwidth}
        \centering
        \frame{\includegraphics[width=1\columnwidth]{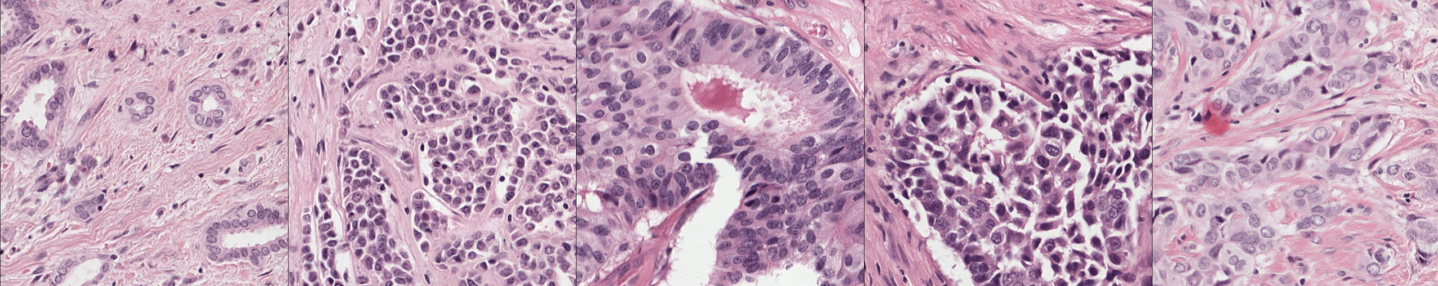}} \\
        \vspace{3pt}
        \frame{\includegraphics[width=1\columnwidth]{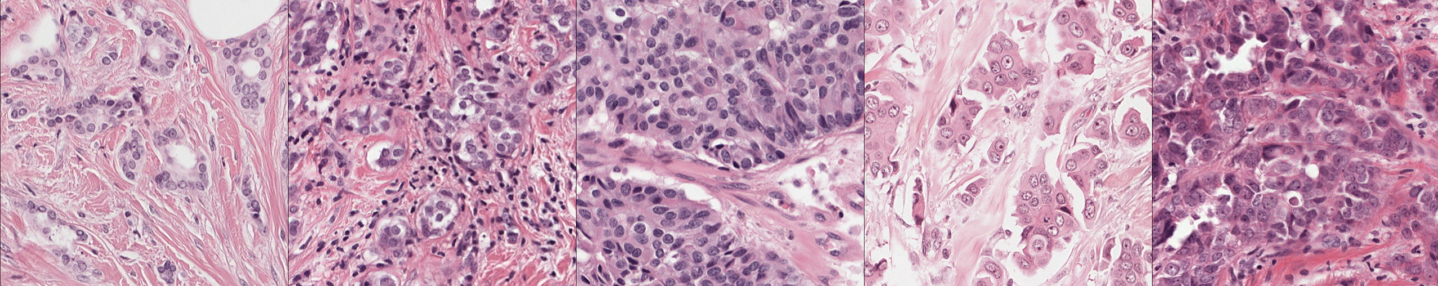}}
        \frame{\includegraphics[width=5pt, height=\columnwidth, angle=270, trim=40 15 10 10, clip]{figures/results/pleo_colorbar.png}} \\
        \vspace{3pt}
        (b)
    \end{minipage} \\
    \vspace{0.02\textwidth}
    \begin{minipage}{.99\textwidth}
        \centering
        \frame{\includegraphics[width=1\columnwidth]{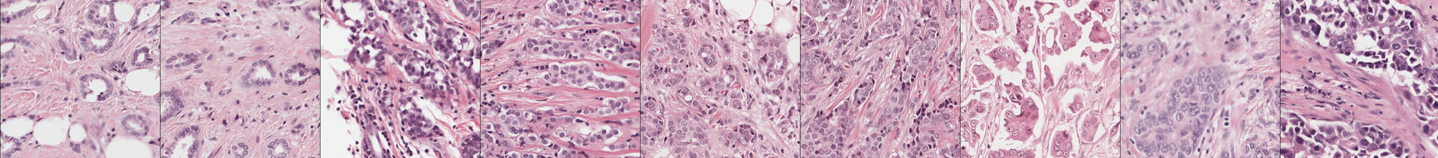}} \\
        \vspace{3pt}
        \frame{\includegraphics[width=1\columnwidth]{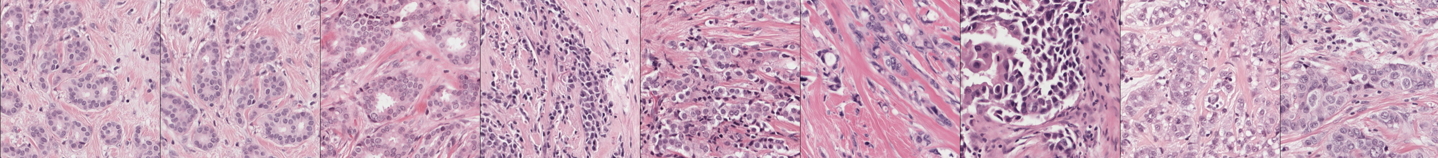}}
        \frame{\includegraphics[width=5pt, height=\columnwidth, angle=270, trim=40 15 10 10, clip]{figures/results/pleo_colorbar.png}} \\
        \vspace{3pt}
        (c)
    \end{minipage}
\caption{Nuclear pleomorphism predictions quantized into different number of categories, sorted from low pleomorphism to high in each category, from left to right. In (a), patches are quantized into three categories which is in line with the traditional three-category classification, whereas in (b) and (c), the patches are quantized into five and nine categories, respectively, to demonstrate the continuity of the nuclear pleomorphism spectrum from the predictions of the \AI. }
\label{figure:results:patch:pleomorphism_grade_spectrum}
\end{figure}

\paragraph{Prediction performance of the \AI on patches.} 
In this section, we analyzed the predictions of the \AI on fixed-sized patches of $512$ $\times$ $512$ pixels randomly cropped at $20 \times$ magnification from $45$ ROIs out of the $13$ evaluation slides in our \ROIdata. 
% \todo{test split etc. was not defined before...}
The output of our \AI is a continuous numerical value ranging between $1-\epsilon$ to $3+\epsilon$, corresponding to the increasing severity of the nuclear pleomorphism.
In order to demonstrate the granularity of our automated approach, we quantized the predictions of the \AI into the three categories as per the guidelines, as well as five and nine additional categories. 
In Figure \ref{figure:results:patch:pleomorphism_grade_spectrum}, we present example patches from the test set for each category, sorted by the quantized predictions of the \AI, with each patch having a higher categorical predicted score than the one preceding it.
While the nuclei in the leftmost patches were closest in appearance to healthy epithelium, the nuclear pleomorphism became gradually more severe with each patch to the right.

\begin{figure}
\centering
    \begin{minipage}{0.47\textwidth}
        \centering
        \begin{minipage}{\columnwidth}
            \centering
            \frame{\includegraphics[width=0.32\columnwidth, trim=200 200 200 200, clip]{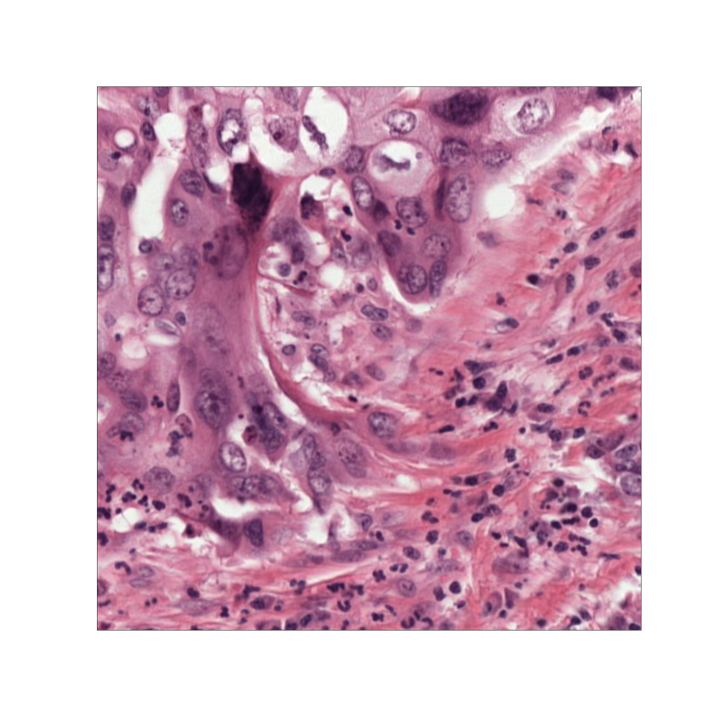}}
            \frame{\includegraphics[width=0.32\columnwidth, trim=200 200 200 200, clip]{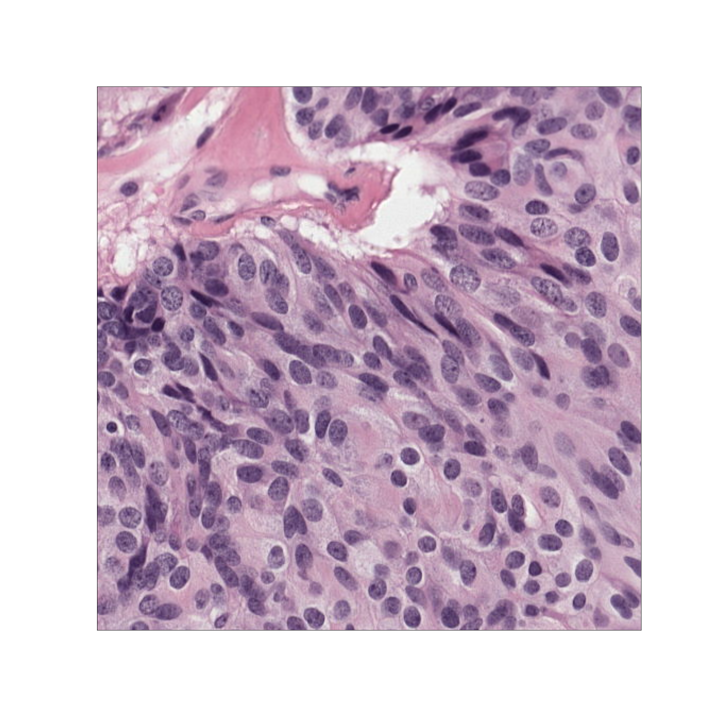}}
            \frame{\includegraphics[width=0.32\columnwidth, trim=200 200 200 200, clip]{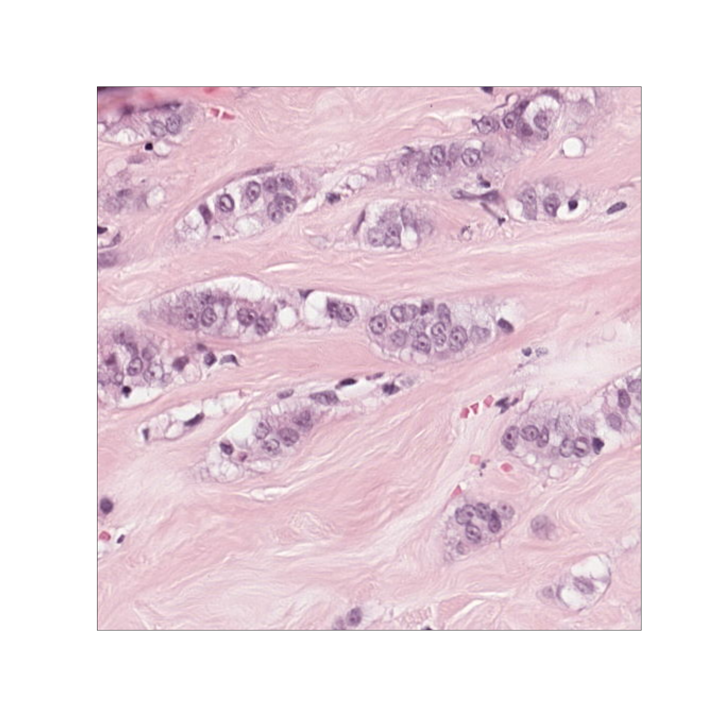}}
        \end{minipage} \\
        \vspace{-1pt}
        \begin{minipage}{\columnwidth}
            \centering
            \frame{\includegraphics[width=0.32\columnwidth, trim=200 200 200 200, clip]{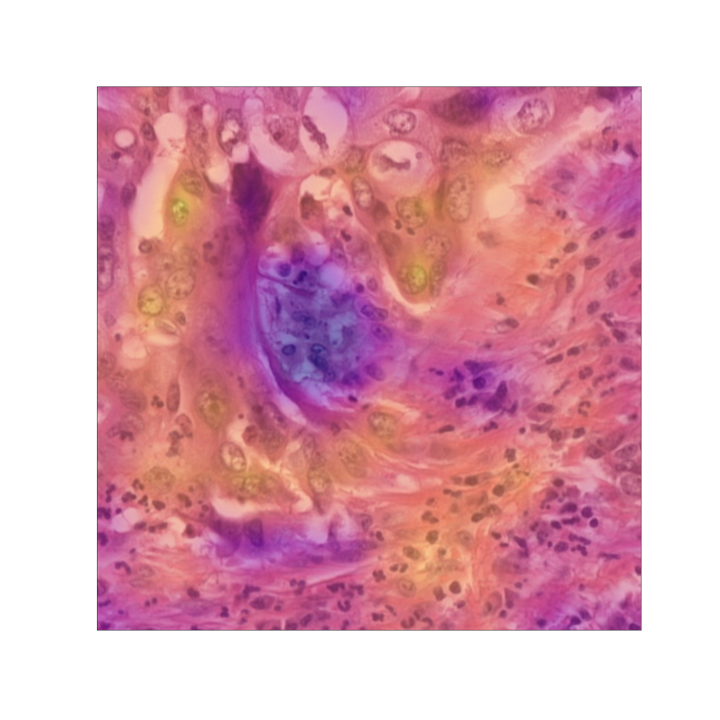}}
            \frame{\includegraphics[width=0.32\columnwidth, trim=200 200 200 200, clip]{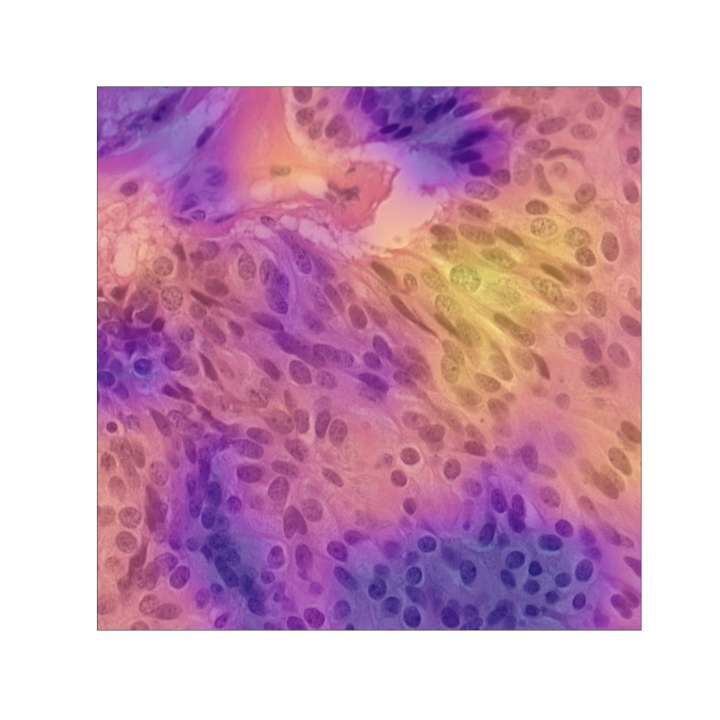}}
            \frame{\includegraphics[width=0.32\columnwidth, trim=200 200 200 200, clip]{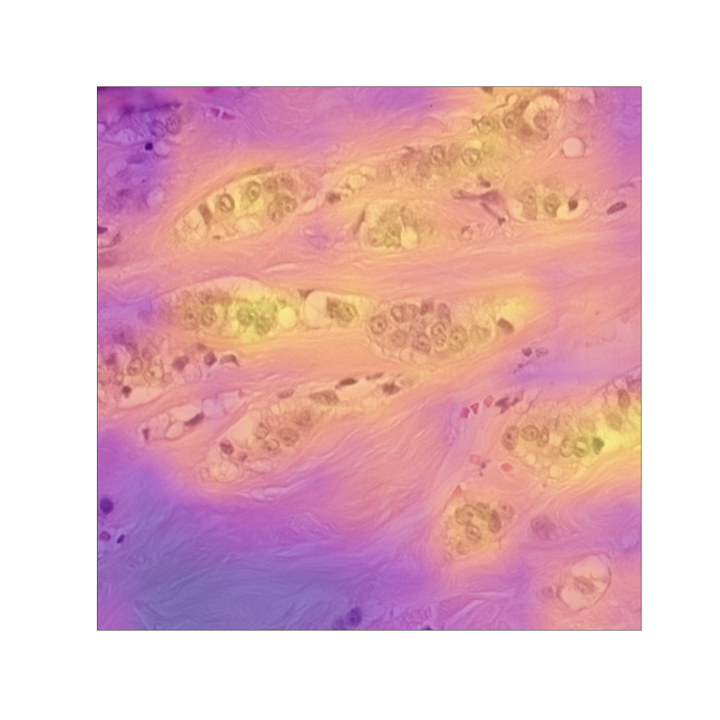}}
            (a)
        \end{minipage}
    \end{minipage}%%%
    \begin{minipage}{0.02\textwidth}
        \vspace{70pt}
        \includegraphics[height=70pt, width=8pt]{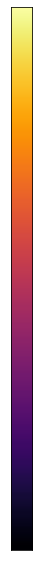}
    \end{minipage}%%%
    \begin{minipage}{0.47\textwidth}
        \centering
        \begin{minipage}{\columnwidth}
            \centering
            \frame{\includegraphics[width=0.32\columnwidth, trim=200 200 200 200, clip]{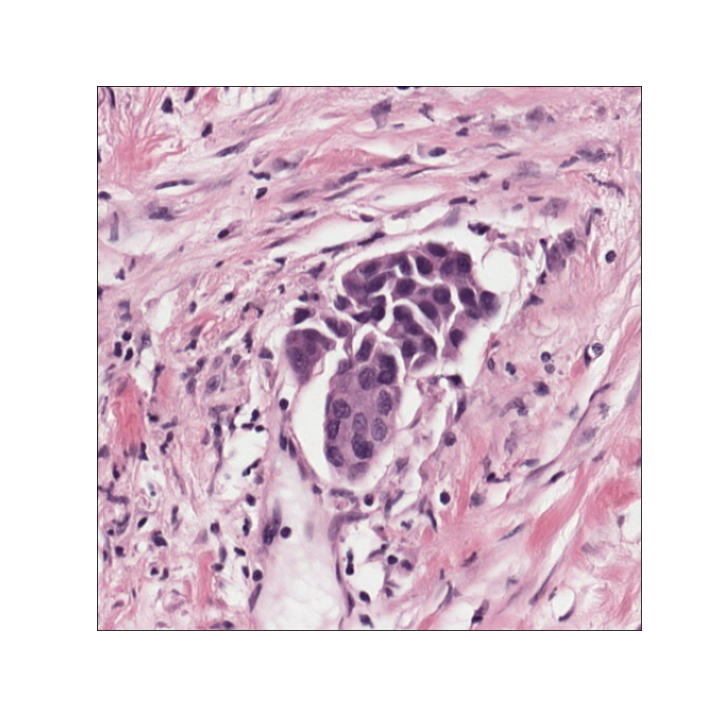}}
            \frame{\includegraphics[width=0.32\columnwidth, trim=200 200 200 200, clip]{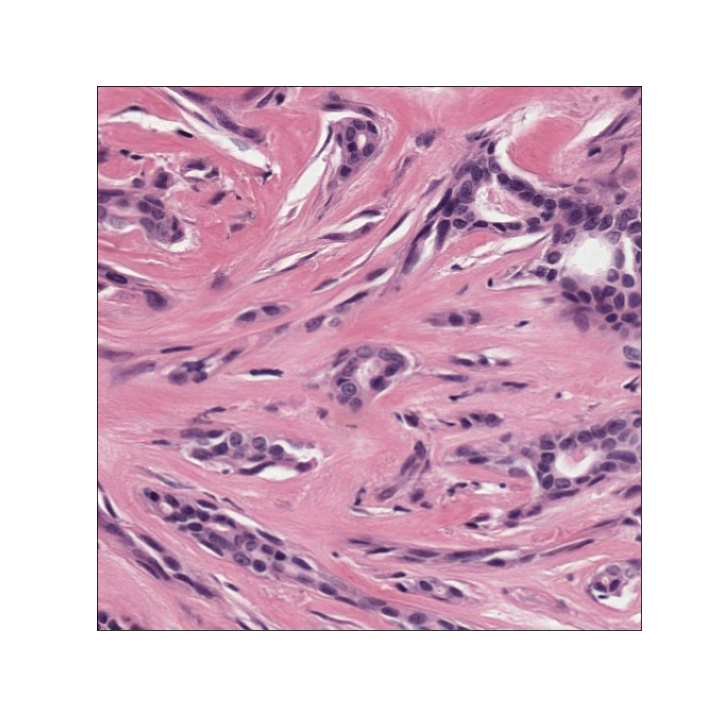}}
            \frame{\includegraphics[width=0.32\columnwidth, trim=200 200 200 200, clip]{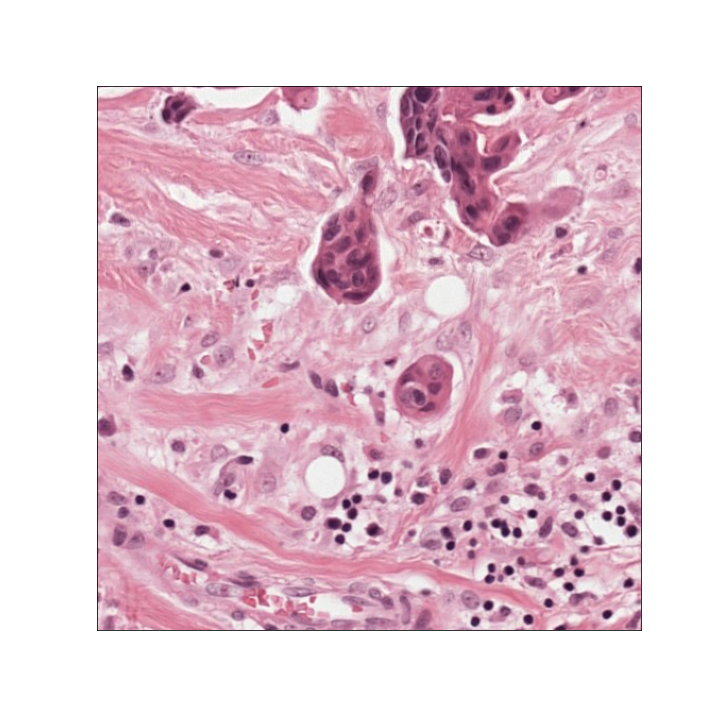}} \\
        \end{minipage} \\
        \vspace{-1pt}
        \begin{minipage}{\columnwidth}
            \centering
            \frame{\includegraphics[width=0.32\columnwidth, trim=200 200 200 200, clip]{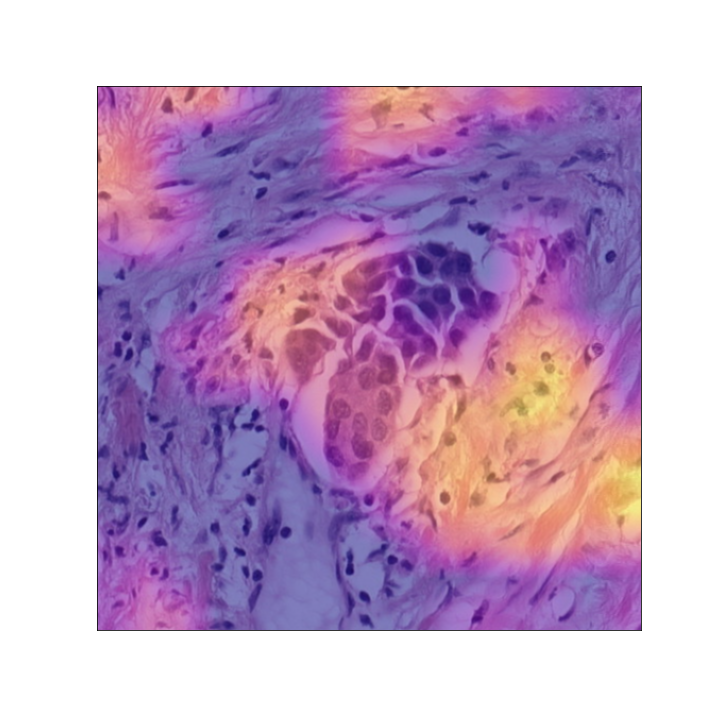}}
            \frame{\includegraphics[width=0.32\columnwidth, trim=200 200 200 200, clip]{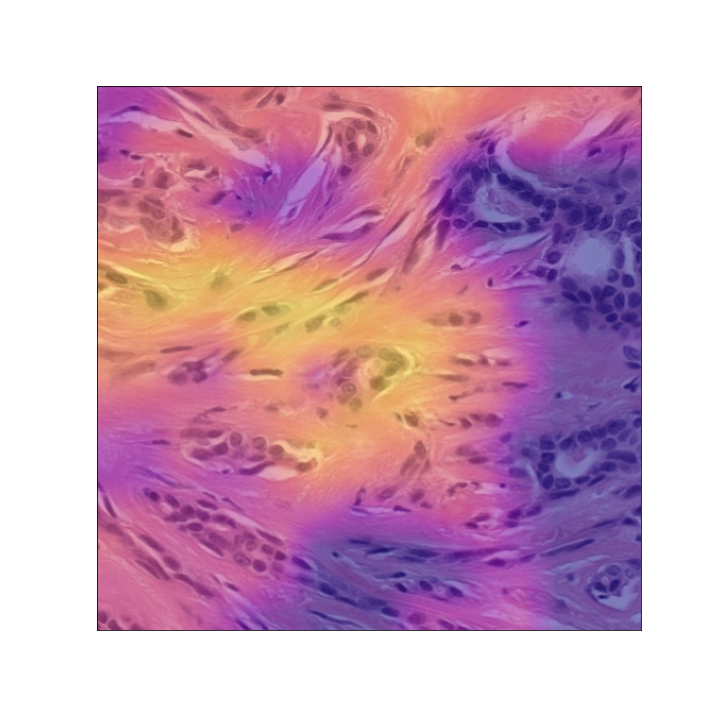}}
            \frame{\includegraphics[width=0.32\columnwidth, trim=200 200 200 200, clip]{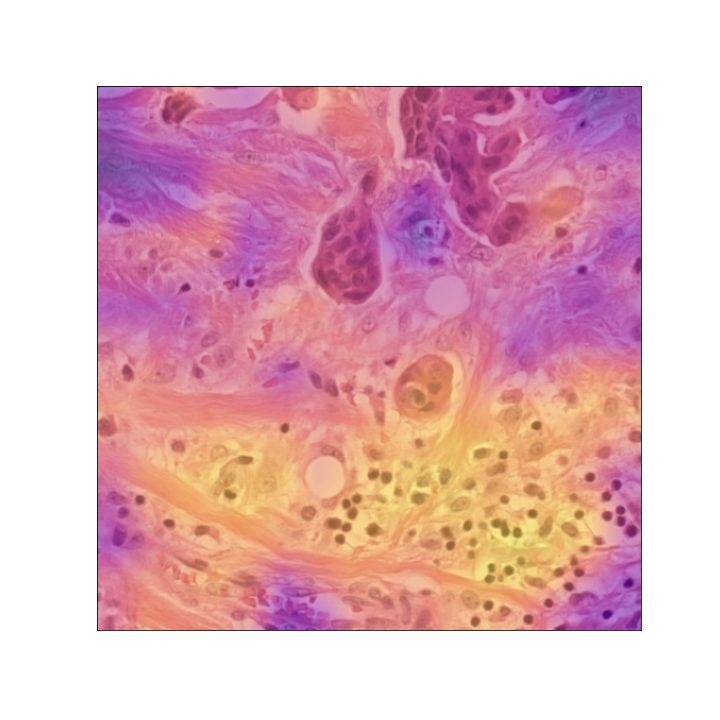}} \\
            (b)
        \end{minipage}
    \end{minipage}
\caption{Grad-CAM saliency outputs of six patches from the evaluation set in the ROI-study. The predictions were close to the reference scores when the \AI focused on the cellular structures (a), and it failed in few cases focusing on other tissue parts (b).}
\label{figure:results:patch:gradcam_outputs}
\end{figure}

Making correct predictions is important, but what is more important is making correct predictions with correct reasoning. 
% \todo{change reasoning; may sound too strong. sth like sanity check etc would be more suitable}
To investigate this, we applied Gradient-weighted Class Activation Mapping (Grad-CAM) \cite{Selv17}, which highlights the areas in an input image patch that contributed most to the prediction output of that patch.
We employed this idea on our \network to see whether the predictions were made by taking the nuclear morphology into account.
In Figure \ref{figure:results:patch:gradcam_outputs}, we present several example patches from the test set in our \ROIdata, visualizing the salient areas determined by the \network.
The patches with predictions close to the reference scores showed strong activation around nuclei.
Similarly, when we inspected the patches where the \AI failed, we observed that it overlooked the nuclear structures and incorporated data from other areas of the tissues.
Our experiments in this section indicate that the \AI predicted similar scores with the pathologists when it focused on the nuclei, failing when focusing on other areas, such as stroma.
% \todo{the part after this is not easy to understand; say sth like multiple horizontally and vertically shifted patches from the same area are extracted to come up with the final scores as will be detailed in the next sections...}
We overcame this problem by processing multiple overlapping patches from the same area with small displacements.
This ensured that the \AI saw a slightly displaced patch and it mostly focused on the same nuclear areas.
As a result, the predictions were, on average, based on the nuclear architectures, as the problem occurred few and far between.

\begin{figure}
\centering
    \begin{minipage}{0.48\textwidth}
        \centering
        \includegraphics[width=\columnwidth, trim=0 380 0 0, clip]{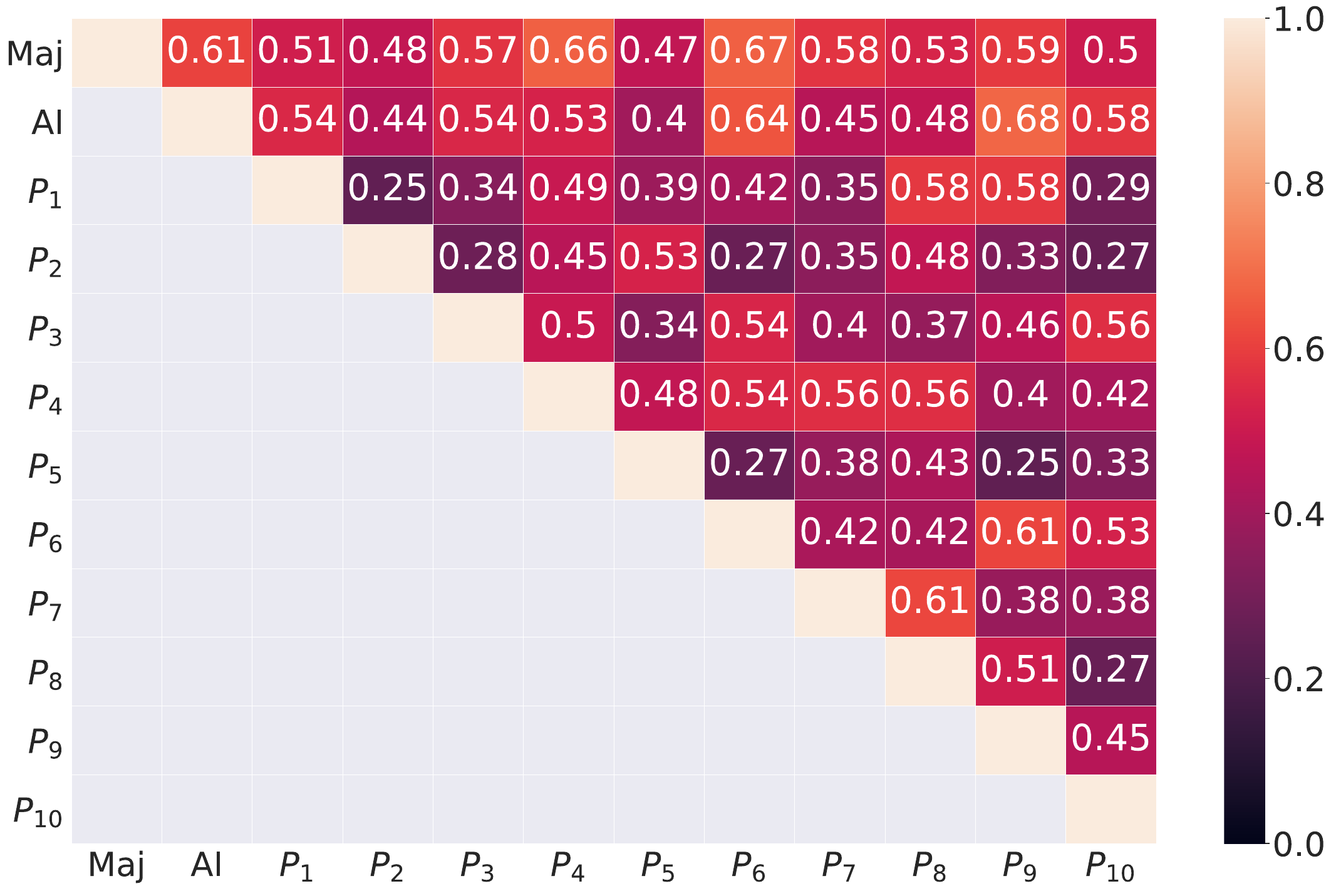} \\
        (a)
    \end{minipage}
    \begin{minipage}{0.48\textwidth}
        \centering
        \includegraphics[width=\columnwidth, trim=0 400 0 0, clip]{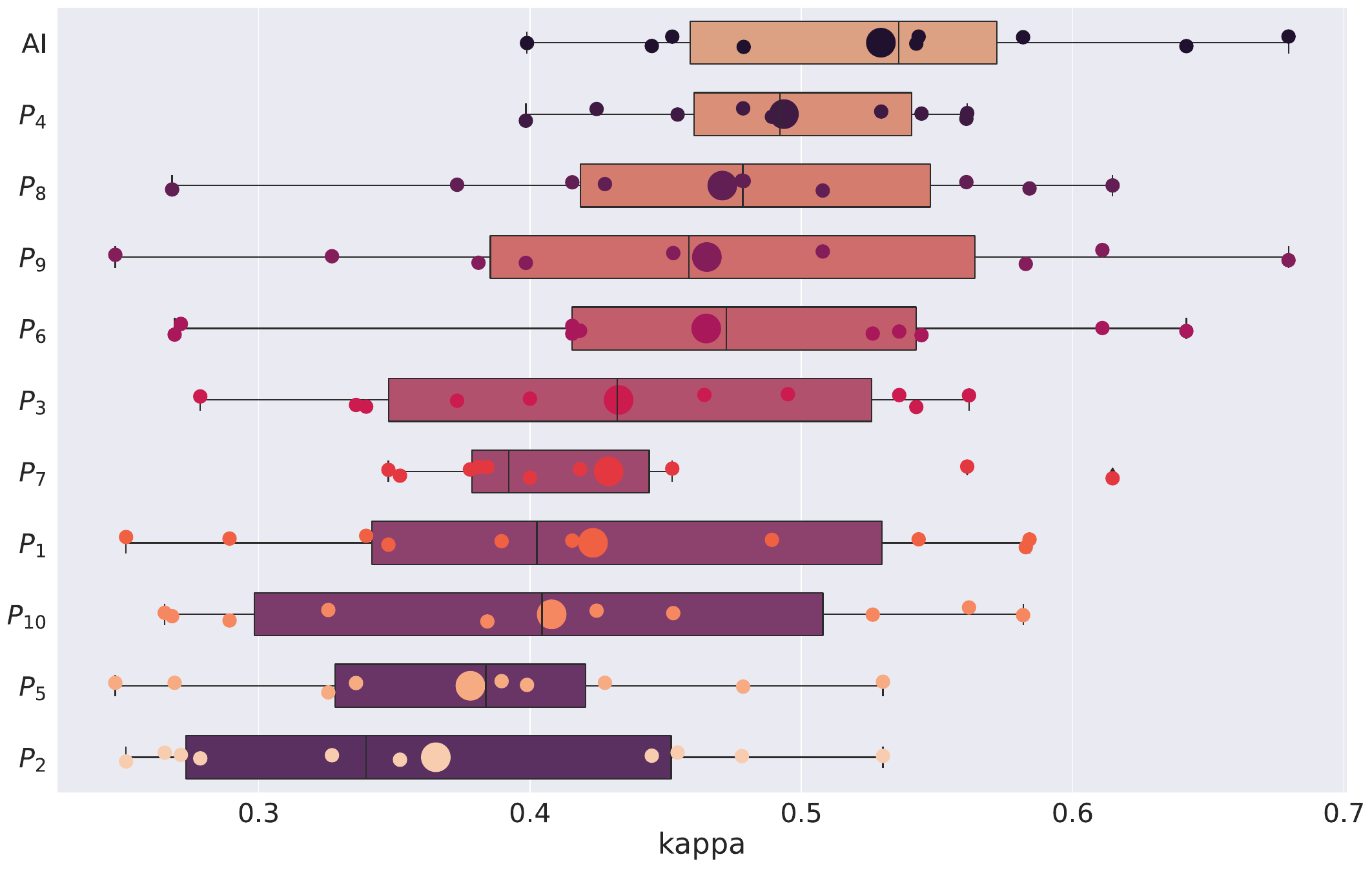}\\
        (b)
    \end{minipage}
\caption{The quadratic kappa scores of the pathologists as well as the \AI on the evaluation set in the ROI-study (45 ROIs). In (a), pairwise kappa scores of the \AI and the pathologists are compared where the best pairwise kappa score was achieved between the \AI and $P_9$. The majority scores, denoted by Maj, had the third best kappa score with the \AI, behind $P_6$ and $P_4$. The pairwise kappa metrics, sorted by the average pairwise scores, (b) shows that the \AI has, on average, the highest agreement out of any pathologists. It also has the highest pairwise kappa score of $0.68$ with $P_9$. In each row, small dots correspond to the kappa scores with the other pathologists/\AI and the large dot denotes the average kappa score of that particular pathologist/\AI with the others.}
\label{fig:results:roi:kappa_comparison}
\end{figure}

\paragraph{Quantitative evaluation of the \AI vs. pathologists.}
The first quantitative comparison between the \AI and the pathologists used the evaluation set of the ROI-study, consisting of $45$ ROIs from $13$ slides. 
ROIs were manually selected to contain as much as possible a uniform degree of pleomorphism.
Each pathologist scored nuclear pleomorphism on these tumor ROIs into one of the three categories, $\{1,2,3\}$. 
We quantized the predicted pleomorphism scores of the \AI for each ROI into the traditional three categories to compare to the scores of the individual pathologists, as well as to the majority vote of their scores (using the provided confidence scores in the case of ties).
We present the kappa scores detailing the agreement comparisons of the pathologists and the \AI in Figure \ref{fig:results:roi:kappa_comparison}.
In this comparison, the scores of a pathologist were not included in the majority voting when that pathologist was compared to the majority scores.
% On the other hand, the scores of all pathologists were used for the majority voting for the comparison between the \AI and the majority scores.
The \AI had a kappa score of $0.61$ with the majority scores (indicated by Maj in the figure), trailing behind only two pathologists, $P_6$ and $P_4$ with kappa scores $0.67$ and $0.66$ with the majority scores, respectively. 
The \AI had the highest average kappa score of $0.53$, followed by the kappa score of $0.49$ by $P_4$.
The \AI on average, showed the highest agreement, and it ranked third behind the two best performing pathologists when compared to the majority scores on the evaluation set in the ROI-study.

\begin{figure}
\centering
    % \vspace{-4pt}
    \begin{minipage}{\textwidth}
        \centering
        % \frame{\includegraphics[height=0.25\columnwidth, trim=40 10 5 10, clip]{figures/results/pleo_colorbar.png}}
        \includegraphics[width=0.30\columnwidth]{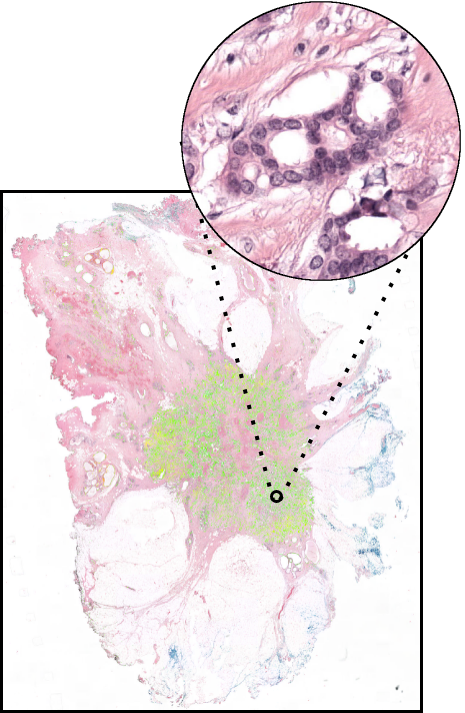}
        \hspace{-10pt}
        % \frame{\includegraphics[height=0.25\columnwidth, trim=40 10 5 10, clip]{figures/results/pleo_colorbar.png}}
        \includegraphics[width=0.30\columnwidth]{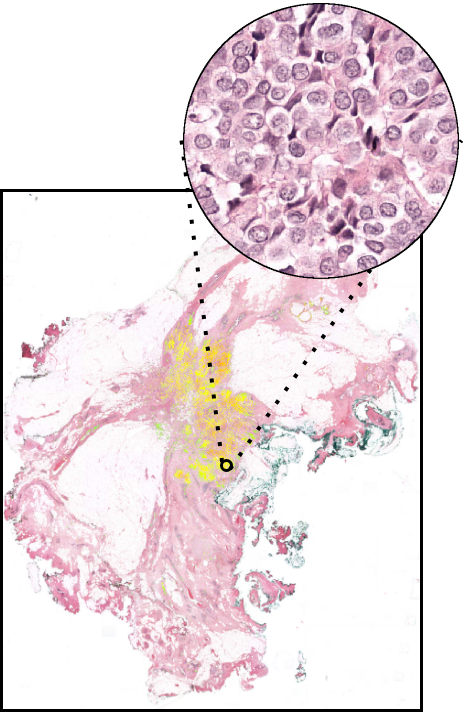}
        \hspace{-10pt}
        % \frame{\includegraphics[height=0.25\columnwidth, trim=40 10 5 10, clip]{figures/results/pleo_colorbar.png}}
        \includegraphics[width=0.30\columnwidth]{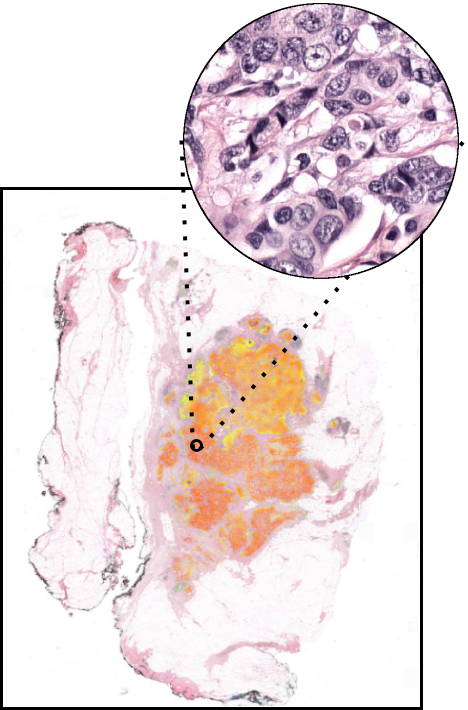}
        % \subcaption{ }
    \end{minipage} \\
    \vspace{5pt}
    \begin{minipage}{\textwidth}
        \centering
        \begin{minipage}{0.65\textwidth}
            \centering
            % \frame{\includegraphics[height=0.50\columnwidth, trim=40 10 5 10, clip]{figures/results/pleo_colorbar.png}}
            \fbox{\includegraphics[width=0.90\columnwidth]{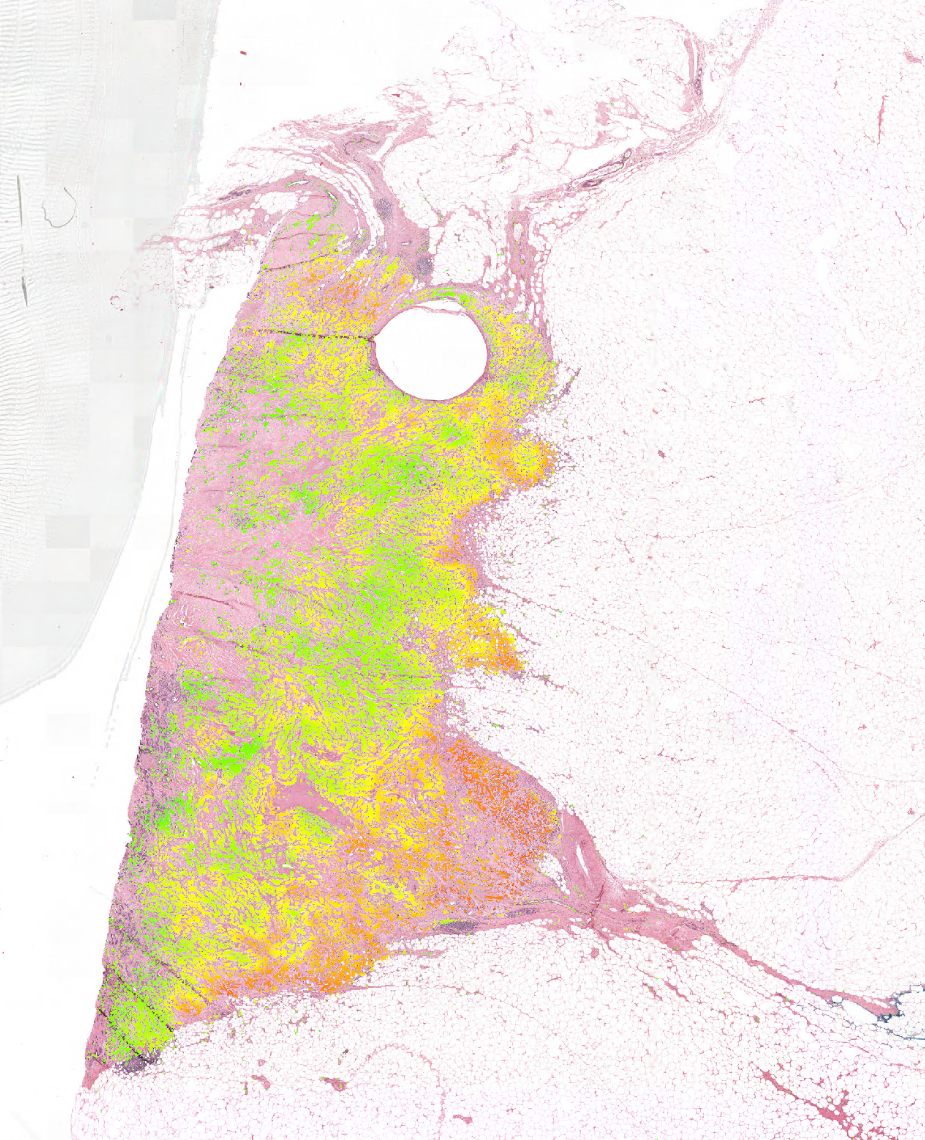}}
        \end{minipage}%%%
        \hspace{-60pt}
        \begin{minipage}{0.30\textwidth}
            \centering
            \includegraphics[width=0.75\columnwidth]{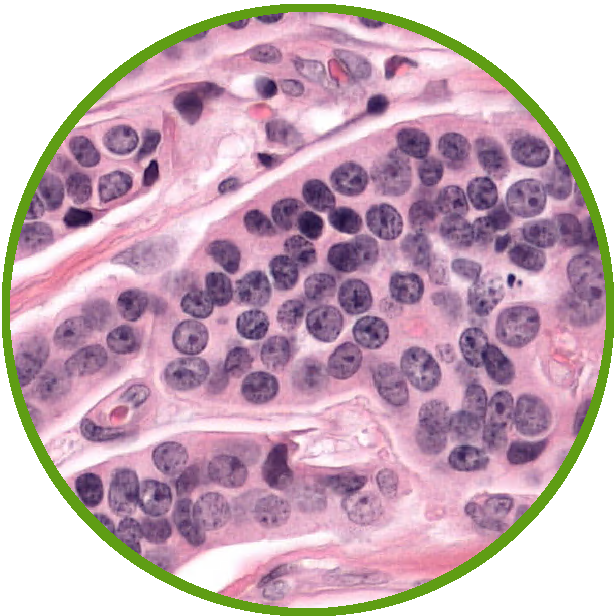} \\
            \includegraphics[width=0.75\columnwidth]{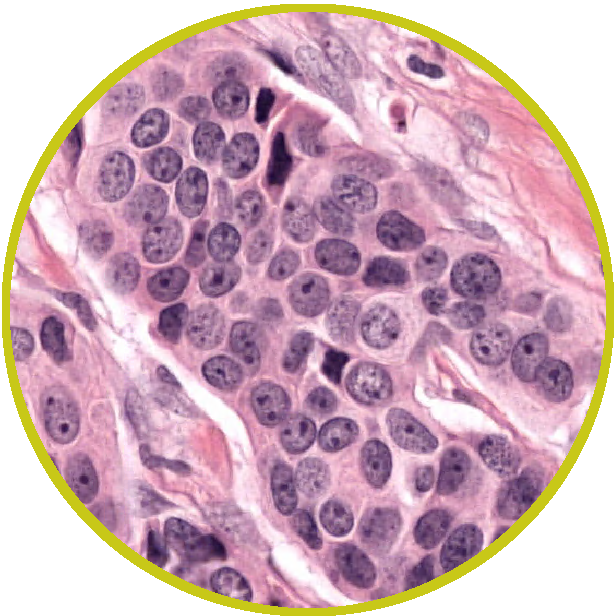} \\
            \includegraphics[width=0.75\columnwidth]{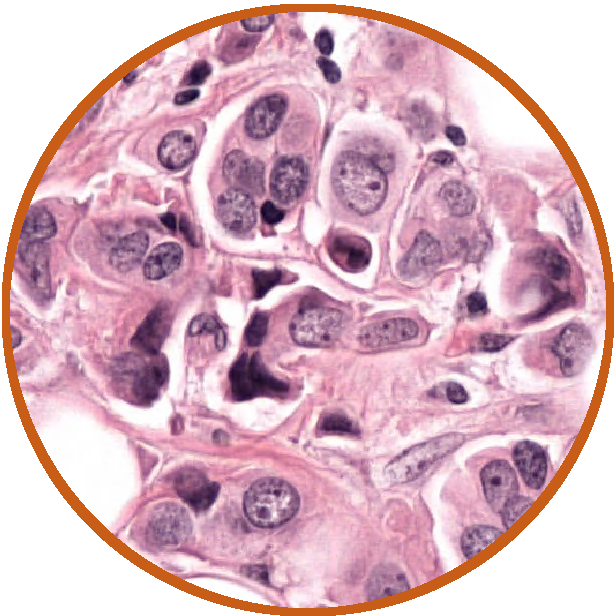}
        \end{minipage}
        % \subcaption{ }
    \end{minipage} \\
    \vspace{5pt}
    \begin{minipage}{\textwidth}
        \centering
        \frame{\includegraphics[width=0.01\columnwidth, height=0.60\columnwidth, angle=270, trim=40 15 10 10, clip]{figures/results/pleo_colorbar.png}}
    \end{minipage}
\caption{Nuclear pleomorphism spectrum of the \AI on four of the $118$ test slides. The color spectrum from green to yellow to red denotes increasing nuclear pleomorphism severity. The deformation in nuclear morphology with the increasing pleomorphism scores demonstrates the capability of the \AI capturing the concept of nuclear pleomorphism.}
\label{fig:results:slide:visual_outputs}
\end{figure}

\paragraph{Real case application by scoring nuclear pleomorphism on whole slide images.}
This slide-level experimental setup was in line with the real-world clinical setting, where an entire slide is assigned a single nuclear pleomorphism score by a pathologist.
The $118$ whole slide images were used for the slide-level evaluation of the \AI compared to the four participating pathologists, denoted by $P_{i}, P_{ii}, P_{iii}, P_{iv}$, who scored nuclear pleomorphism on the slides into one of the three categories.
In contrast to the ROI-level study in which ROIs were selected to have homogeneous pleomorphism, tumor in whole slide images generally contained larger heterogeneity with more diverse nuclei morphology.
We present the visual pleomorphism spectrum of the (non-quantized) predictions of the \AI on four example slides from the slide-study in Figure \ref{fig:results:slide:visual_outputs}.

\begin{figure}
\centering
    \begin{minipage}{0.6\textwidth}
        \begin{minipage}{0.5\columnwidth}
            \centering
            \includegraphics[width=\columnwidth]{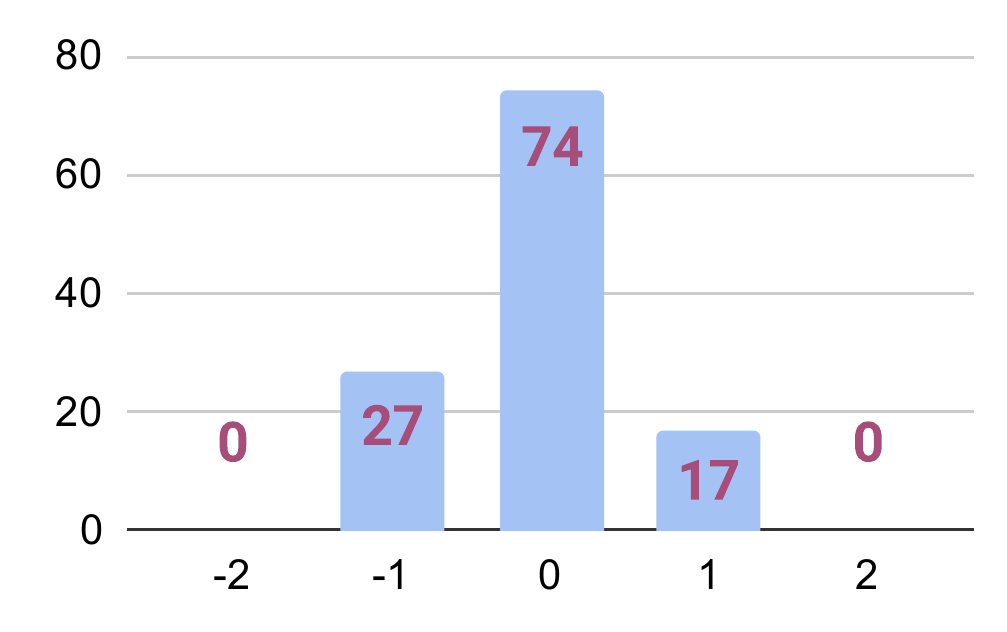}
            \subcaption{AI vs. $P_{i}$}\label{fig:results:slide:kappa_scores_difference1}
        \end{minipage}%%%
        \begin{minipage}{0.5\columnwidth}
            \centering
            \includegraphics[width=\columnwidth]{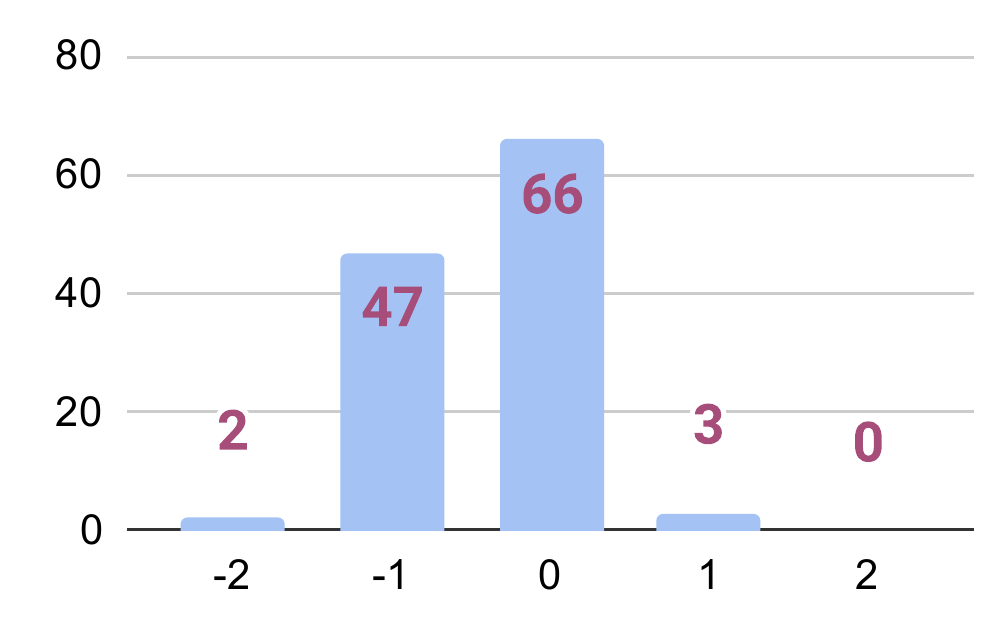}
            \subcaption{AI vs. $P_{ii}$}\label{fig:results:slide:kappa_scores_difference2}
        \end{minipage}
        \\
        \begin{minipage}{0.5\columnwidth}
            \centering
            \includegraphics[width=\columnwidth]{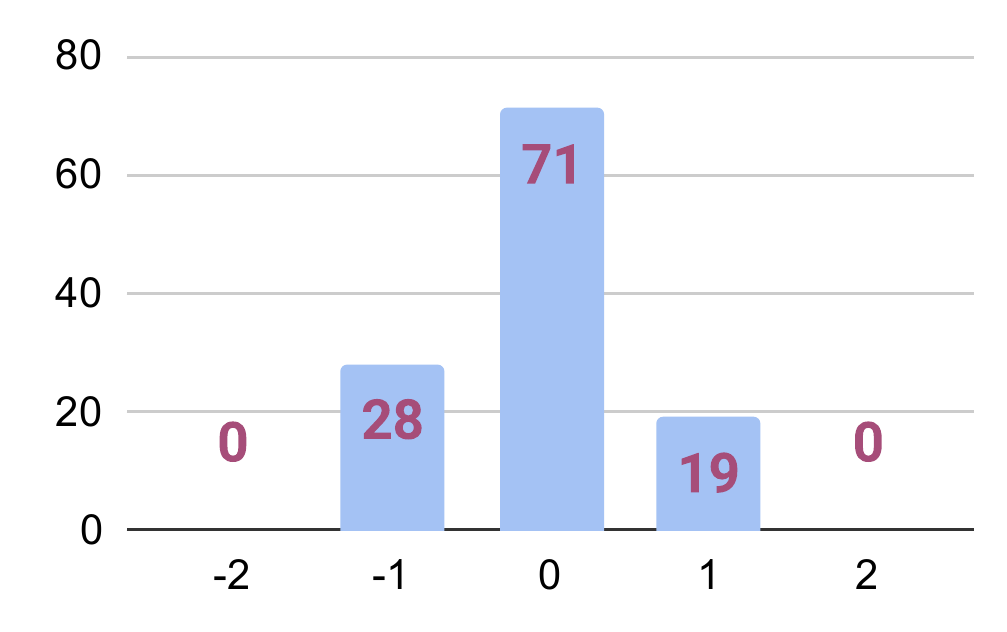}
            \subcaption{AI vs. $P_{iii}$}\label{fig:results:slide:kappa_scores_difference3}
        \end{minipage}%%%
        \begin{minipage}{0.5\columnwidth}
            \centering
            \includegraphics[width=\columnwidth]{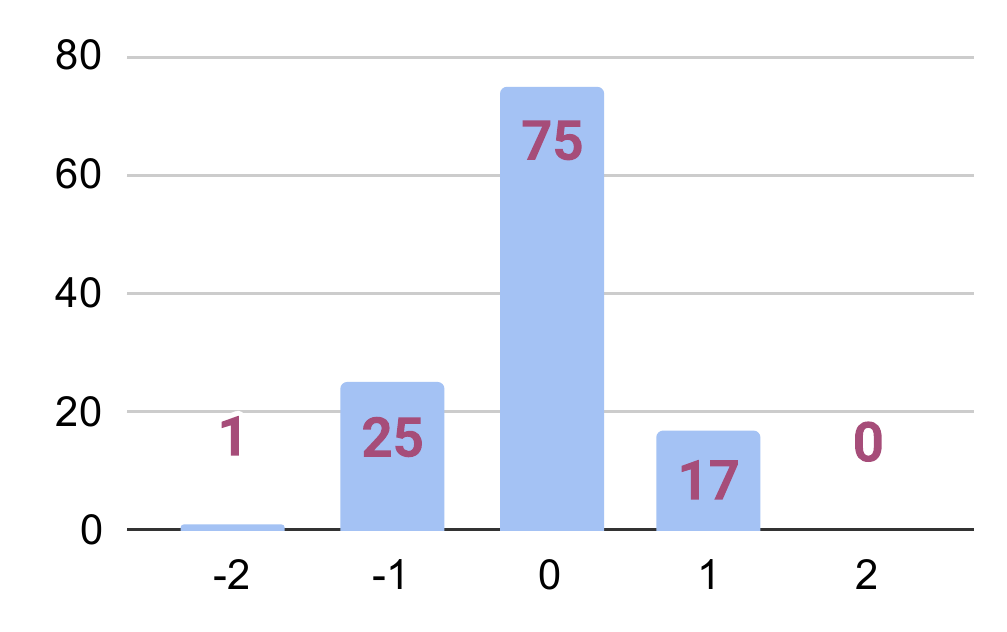}
            \subcaption{AI vs. $P_{iv}$}\label{fig:results:slide:kappa_scores_difference4}
        \end{minipage}
    \end{minipage}%%%
    \hfill\vline\hfill
    % \hspace{2pt}
    \begin{minipage}{0.35\textwidth}
    \centering
        \begin{minipage}{0.85\columnwidth}
            \centering
            \includegraphics[width=\columnwidth]{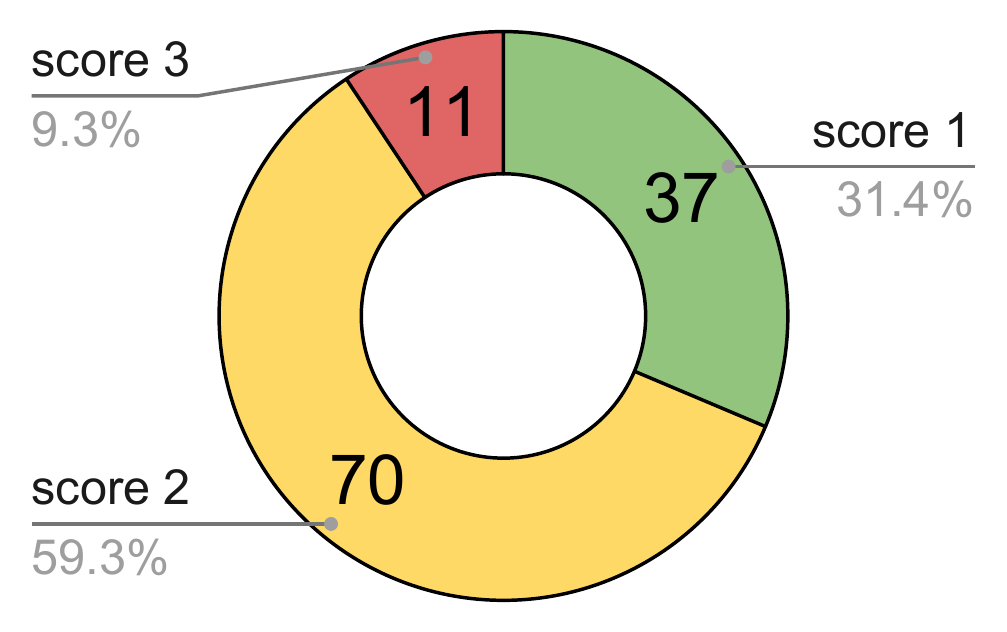}
            \subcaption{}\label{fig:results:slide:AI_prediction_distribution}
        \end{minipage}
        \\
        \begin{minipage}{\columnwidth}
            \centering
            \includegraphics[width=\columnwidth, trim=0 330 0 0, clip]{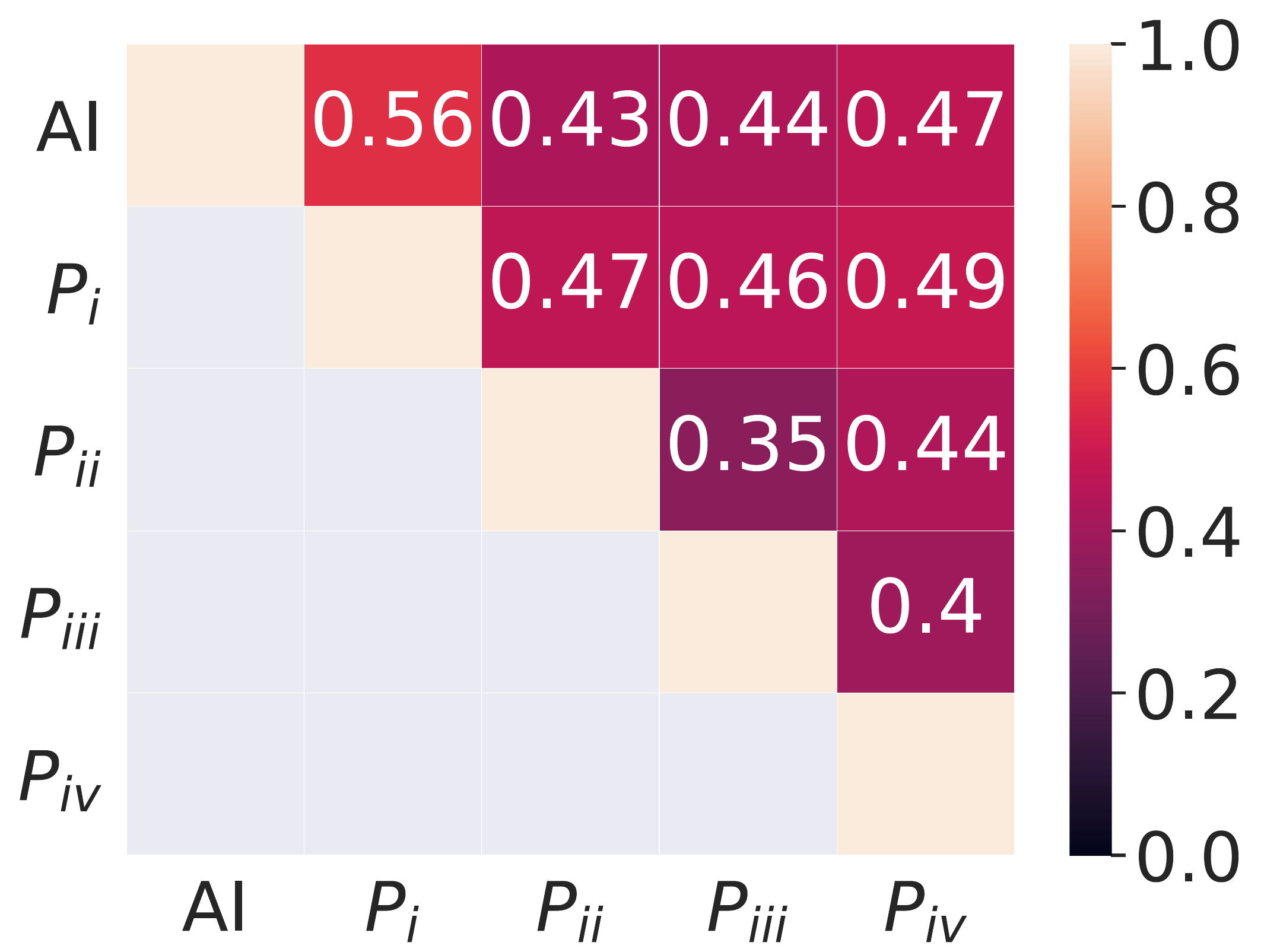}
            \subcaption{}\label{fig:results:slide:kappa_scores_comparison}
        \end{minipage}
    \end{minipage}
\caption{The evaluation of the performance of the \AI on $118$ evaluation slides in the slide-study. The scores of the \AI are quantized into three categories. The differences in scores of the \AI from the scores given by the four pathologists are presented through (a) - (d). The distribution of the \AI scores is shown in (e). The kappa scores of the \AI and the four pathologists (f) show that the \AI and $P_i$ has the highest agreement, and the \AI, on average pairwise kappa scores, is only second to $P_i$.}
\label{fig:results:slide:kappa_scores}
\end{figure}

For the quantitative analysis of the \AI compared to the pathologists, we quantized the pleomorphism predictions of the slides into one of the three categories.
Figures \ref{fig:results:slide:kappa_scores_difference1} - \ref{fig:results:slide:kappa_scores_difference4} present the score difference of the \AI compared to the pathologists.
The \AI achieved a nuclear pleomorphism score equal to pathologists $P_{i}, P_{ii}, P_{iii}, P_{iv}$ in $74, 66, 71$ and $75$ slides, respectively.
The \AI had a score difference of $\pm1$ on the rest of the slides, except for the two slides in comparison to $P_{ii}$ and one slide to $P_{iv}$ where the pathologists scored $3$ for the slides whereas the \AI scored $1$.
For these slides, the scores of the other pathologists were ($1, 2, 1$), ($2, 2, 2$) and ($2, 2, 2$). 
Additionally, compared to the score differences of the pathologists, as seen in Figure \ref{fig:data:slide:pathologists_grade_stats}, the \AI had the most matching scores in $75$ slides with $P_{iv}$ as well as the highest matching scores, on average.
Looking at the score distribution of the \AI, we observe that the \AI predicted the score $1$ in $37$ slides, score $2$ in $70$ slides and score $3$ in $11$ slides, as shown in Figure \ref{fig:results:slide:AI_prediction_distribution}.
Finally, we present the kappa scores of the \AI as well as the individual pathologists in Figure \ref{fig:results:slide:kappa_scores_comparison}. 
The highest kappa score of $0.56$ was achieved between the \AI and $P_{i}$. 
The \AI had kappa scores of $0.43, 0.44$ and $0.47$ with the rest of the pathologists, $P_{ii}, P_{iii}, P_{iv}$, respectively. 
Overall, the average kappa score for the \AI was $0.475$ and it was only second to $P_{i}$ with $0.497$.
$P_{i}$ was also the participant with the overall highest average agreement.
The results indicate that $P_i$ agreed the most with the \AI, more so than any other pathologists.
Our findings in this slide-level study were consistent with our results in the previous experiments with the \AI consistently ranking high in agreement with the best performing pathologists.
Additionally, qualitative analysis of the predictions on whole slide images demonstrated that the \AI learned to score the full spectrum of nuclear pleomorphism based on morphological changes in tumor cells.

\section{Discussion}
In this paper, we proposed a novel method for automated scoring of nuclear pleomorphism in breast cancer as a continuous feature rather than a discrete classification.
Our \AI was composed of two stages.
In the first stage, an \celldetector was used to locate tumor cells in whole slide images.
In the second stage, a \network scored nuclear pleomorphism on the tumor, considering the tumor pleomorphism as a spectrum.
Our results showed that the \AI consistently achieved top-level performance similar to the best performing pathologists throughout our experiments in ROI and slide-level studies.

In our experiments, we quantized the patch-level predictions of the \AI into arbitrary number of categories to showcase its flexibility of demonstrating the continuity of the pleomorphism in tumor in greater granularity than the traditional three-category classification.
In the few cases, the predictions failed to capture the pleomorphism in tumor when the \AI did not focus on areas with tumor cells. 
We discovered that this shortcoming could easily be bypassed by processing multiple overlapping patches from the same area and average pooling the predictions, as the occurrence of the problem was few and far between.
We also investigated the added value of normal epithelium as reference, following the workflow of the pathologists in routine practice.
The normal ROIs were less diverse compared to tumor ROIs, due to the slides containing only a very small number of fairly small areas with normal cells, and the inherent modest variation in normal structures.
In addition, due to the nature of normal epithelium, normal ROIs contained, on average, a larger ratio of stroma and ductal structures to cells.
Therefore, we do not rule out the possibility that the \AI might have focused on such structures more and failed to capture the nuclear morphology of normal cells.
We leave the further investigation of this learning setup for future work.
However, we proved that our automated approach could already learn the entire spectrum of tumor pleomorphism without additional knowledge of the normal cell morphology through multiple experiments.

In the ROI-study, the \AI achieved a higher agreement with the consensus of pathologists and highest pairwise kappa scores versus any individual pathologist. 
This is an important outcome as it suggests that our approach is the best out of the participating pathologists at scoring pleomorphism on homogeneous tumor regions.
The results of our experiments in the slide-study suggested a similar outcome with the \AI achieving the second best average kappa score among four pathologists. 
In addition, the best performing pathologist with the highest average kappa score agreed the most with the \AI by a large margin.
In very few cases, where the \celldetector captured the tumor in regions with Ductal Carcinoma in Situ (DCIS), the \network scored pleomorphism on those regions, too.
We argue that the performance of the \AI was slightly hampered by such false predictions causing the overall score of the slides to be quantized one score off from the predictions of the \network for the tumor actually suggested.
% maybe an example image with the predictions on the image with DCIS was scored; changing the overall score of the slide.
This is also visible in the Figure \ref{fig:results:slide:AI_prediction_distribution}, as the predictions of the \AI, on average, were slightly lower than that of the pathologists, as presented in Figure \ref{fig:data:roi:grade_distribution}.
This subsidiary issue could be mitigated by enriching the training set of the \celldetector with more examples from DCIS cases.

In this work, we proposed a fully automated deep learning methodology for the generalized version of the traditional three-category classification of nuclear pleomorphism in breast cancer, formulated as a continuous spectrum of the entire tumor morphology.
We validated the performance of the \AI through multiple qualitative and quantitative experiments, and showed that our approach could reach agreement with a reference standard similar to that of the best performing pathologists.
% We argue that our findings raise an important question whether nuclear pleomorphism should be considered more of a spectrum of continuous change in tumor morphology rather than a three-category classification.
The output of our approach is the full spectrum of tumor pleomorphism, which is easy to analyse, interpret and reproduce.
The full pleomorphism spectrum is highlighted for the pathologists to analyse the tumor morphology without having to zoom-in to investigate individual cells in multiple tumor regions.
Similarly, overall score of a slide could easily be interpreted by the pathologists from the distribution of the pleomorphism spectrum visualized on the slide.
Therefore, our work could be a very efficient tool for pathologists in their daily workflow for nuclear pleomorphism scoring, also leveraging the problem of subjective interpretation due to the deterministic nature of the automated predictions, and therefore improving concordance among pathologists.
Our future work will include such a study in which we will investigate whether the addition of the \AI as a support system could improve the agreement between pathologists.
Another future work will include the investigation of the prognostic value of the automated scoring of nuclear pleomorphism spectrum in breast histopathology.
% \todo{comment from Mark: Another key point to consider is regional variation within a tumor. If a tumor shows the same morphology throughout, then evaluation is relatively simple and review is quick by a pathologist. The problem is when there is heterogeneity, how can we capture that feature, quantify it and see if it has meaning. This is also a struggle for a pathologist, as it forces us to choose one number when we see this range of changes. Accordingly, we go with the worst area. In that regard, has your analysis looked at some way of defining the worse area per tumor? The most aggressive part may drive prognosis and direct management.}

\section{Methods}
\label{sec:methodology}
\paragraph{ROI-study for the training and validation of the \AI.}
We collected H\&E stained whole slide images of breast cancer resections, denoted as $\mMS{W}=\{\mSS{W}_1, \mSS{W}_2, \ldots, \mSS{W}_S\}$, with a total number of $S=39$ slides from two cohorts, which we refer to as cohort A and cohort B.
The slides in cohort A were scanned on a 3DHistech P1000 scanner at \SI{0.25}{\micro\metre}/pixel, and the slides in cohort B were scanned on a 3DHistech Pannoramic 250 Flash II scanner at the same spatial resolution of \SI{0.25}{\micro\metre}/pixel.
From the cohort A; we made a balanced selection of cases with respect to the nuclear pleomorphism scores.
Invasive Ductal Carcinoma (IDC) was the most prevalent tumor type, seen in 24 out of the 31 slides, resembling the distribution of breast cancer patients in clinical practice.
From the cohort B; we included 8 additional cases that covered different histological subtypes with severely aberrant morphology from a triple negative breast cancer cohort. 
As a result, pleomorphism of the tumor with score 3 in cohort A was, on average, less severe than the tumor with score 3 in cohort B.
Overall, the slides were selected to cover the entire range of nuclear pleomorphism from a large spectrum of tumor morphology in breast cancer pathology, as presented in Table \ref{tab:data:tumor_type}. 

In routine clinical practice, normal epithelium in a whole slide image, when present, is a useful reference point for pathologists to determine the degree of pleomorphism of the tumor.
In the ROI-study, we present pairs of ROIs from tumor and normal epithelium to the pathologists to perform nuclear pleomorphism scoring similar to the routine practice.
We manually selected one or more tumor ROIs from each slide, $\mMR{\hat{R}}_{s} = \{ \hat{R}_{s1}, \hat{R}_{s2}, \dots, \hat{R}_{s{t_s}} \}$, ensuring score homogeneity of tumor cells within each ROI where $t_s$ denotes the number of selected tumor regions from slide $\mSS{W}_s$. 
Overall, we selected $T=125$ such regions from $39$ slides, $\mMS{\hat{R}} = \{\mMR{\hat{R}}_1, \mMR{\hat{R}}_2, \dots \mMR{\hat{R}}_T\} \subset \mMS{W}$. 
Additionally, we manually selected at least one ROI, $\mMR{\bar{R}}_s = \{ \bar{R}_{s1}, \bar{R}_{s2}, \dots, \bar{R}_{s{n_s}} \}$ from normal epithelium (normal ROI), where $n_s$ denotes the number of such regions selected from the slide $\mSS{W}_s$. 
As few slides contained little to no large enough areas with normal epithelium, it was not possible to select at least one normal ROI from every slide.
In total, there were $59$ normal ROIs that we could select from the set of slides, $\mMS{\bar{R}} = \{ \mMR{\bar{R}}_1, \mMR{\bar{R}}_2, \dots \mMR{\bar{R}}_N\} \subset \mMS{W}$. 
Ten out of the $39$ slides did not have large enough region with normal cells. 
Therefore, for the whole slide images without normal epithelium, we retrieved $10$ additional slides from the same patient to acquire at least one normal ROI.
As a result, the total number of selected normal ROIs increased to $N=79$ with the selection of additional $20$ normal ROIs from the extra set of slides.
Following this selection procedure, each ROI was cropped around their center point to a standard square area of around $0.38~mm^2$ at $40\times$ magnification. 
Subsequently, a tumor ROI was paired with a normal ROI from the same patient to form a query pair, $ Q_{s,jk} = \{\hat{R}_{sj}, \bar{R}_{sk}\}$, $j = \{1,\ldots,t_s\}$, $k \in \{1,\ldots,n_s\}$. 
This process is repeated for all $125$ tumor ROIs in this data set. 
It has to be noted that each query had a unique tumor ROI, but a normal ROI could be paired with more than one tumor ROI due to the smaller number of normal ROIs vs tumor ROIs, $79$ vs $125$.

\begin{figure}
\centering
\begin{minipage}{0.475\textwidth}
    \centering
    \begin{minipage}{\columnwidth}
        \begin{minipage}{0.48\columnwidth}
            \centering
            \frame{\includegraphics[width=\columnwidth]{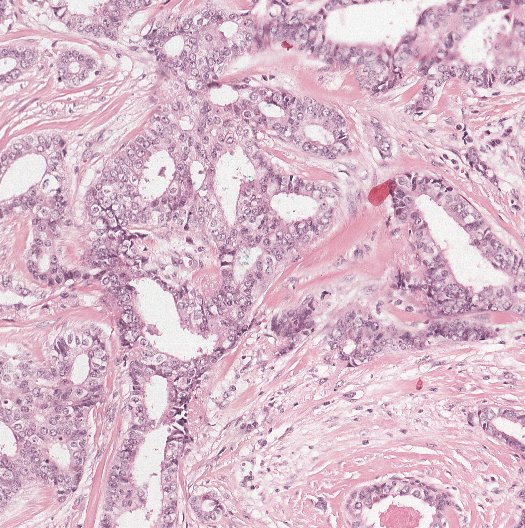}}
        \end{minipage}%%%
        \hspace{5pt}
        \begin{minipage}{0.48\columnwidth}
            \centering
            \frame{\includegraphics[width=\columnwidth]{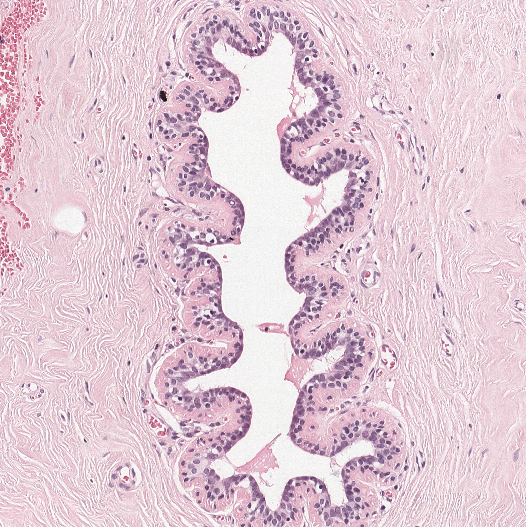}}
        \end{minipage}
        \\
        \begin{minipage}{0.48\columnwidth}
            \centering
            \includegraphics[height=0.50\linewidth]{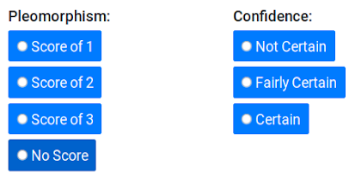}
        \end{minipage}%%%
        \hspace{5pt}
        \begin{minipage}{0.48\columnwidth}
            \centering
            \includegraphics[height=0.50\linewidth]{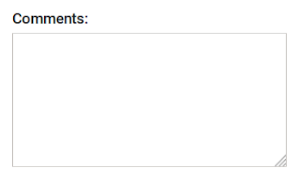}
        \end{minipage}
    \end{minipage}
    \subcaption{An example query from the web platform of the ROI-study with a pair of tumor and normal ROIs from the same patient. Ten pathologists are invited to score nuclear pleomorphism and provide confidence scores for such $125$ queries from $39$ whole slide images.}\label{fig:data:roi:query}
\end{minipage}
\hspace{5pt}
\begin{minipage}{0.495\textwidth}
    \centering
    \begin{minipage}[t]{\columnwidth}
        \centering
        \includegraphics[height=0.80\columnwidth]{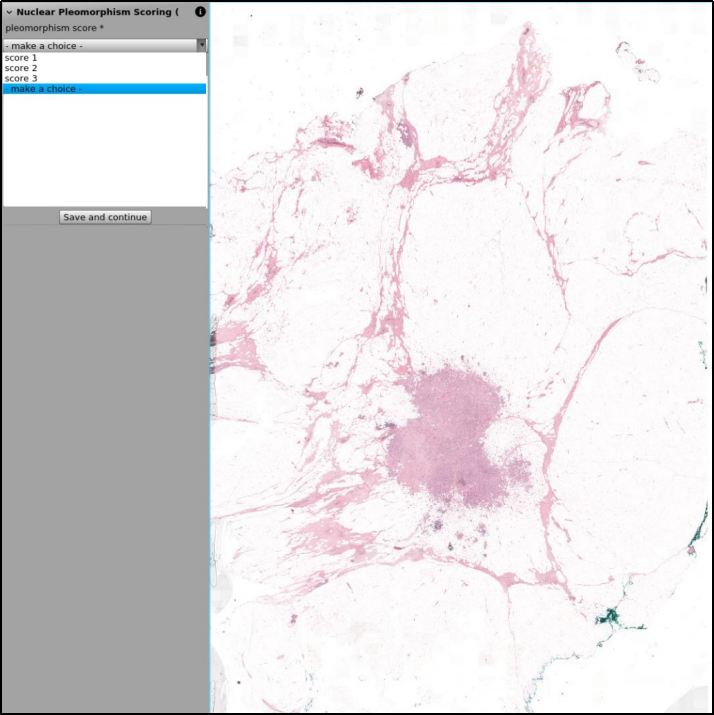}
        \subcaption{An example query from the web platform for the slide-study. Four pathologists are invited to score nuclear pleomorphism for $118$ whole slide images.}\label{fig:data:slide:query}
    \end{minipage}
\end{minipage}
\caption{The web platforms utilized for the (a) ROI-study and (b) slide-study.}
\label{fig:data:studies}
\end{figure}

\begin{figure}
    \centering
    \begin{minipage}[][][b]{0.03\textwidth}
         $P_1$ \\ $P_2$ \\ $P_3$ \\ $P_4$ \\ $P_5$ \\ $P_6$ \\ $P_7$ \\ $P_8$ \\ $P_9$ \\ $P_{10}$ \vspace{30pt}
    \end{minipage}%%%
    \begin{minipage}{0.96\textwidth}
        \includegraphics[height=0.31\textwidth, width=\textwidth]{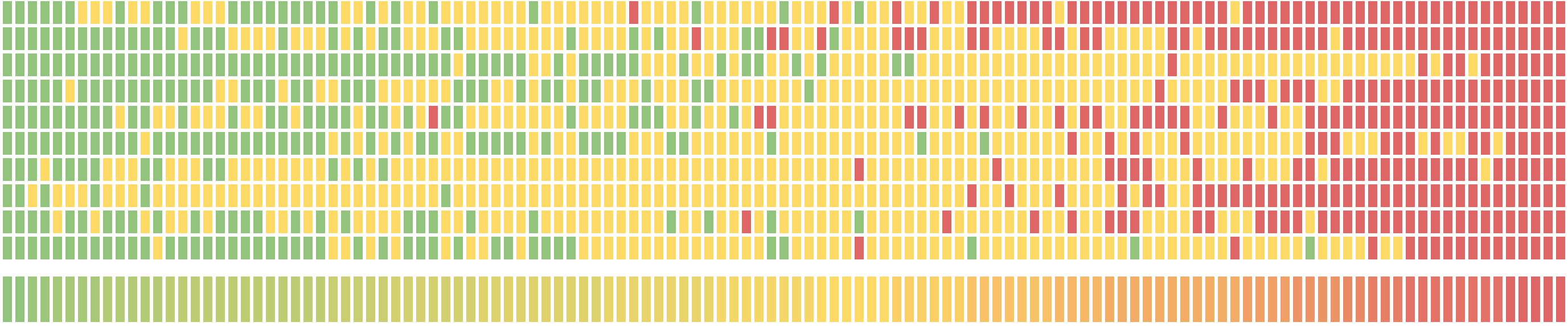}
    \end{minipage}
    \caption{Distribution of the nuclear pleomorphism scores of the pathologists and of the average (reference) scores of the pathologists in the ROI-study. Each row corresponds to a pathologist, except for the last row which corresponds to the average scores of the pathologists which we use as reference scores in our learning setup. Each column is one of the $125$ queries in the ROI-study. The colors green, yellow and red correspond to the scores of $1$, $2$ and $3$ of the pathologists, respectively. The queries are sorted by the average scores, from left to right in increasing order, denoted by the color hues of green, yellow and red, corresponding to the increasing severity of the pleomorphism. }
    \label{fig:data:roi:grade_distribution}
\end{figure}

\begin{figure}
    \centering
    \includegraphics[width=\textwidth]{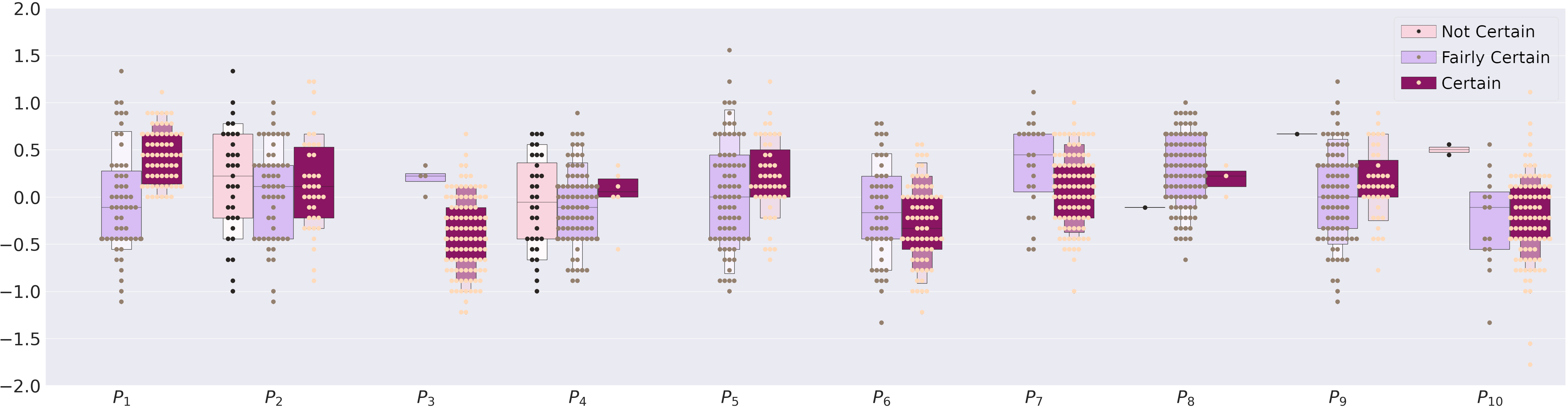}
    \caption{Nuclear pleomorphism score distribution of the pathologists in the ROI-study with respect to their difference from the reference scores, categorized by pathologist confidence. Each box plot and the points within correspond to the score differences of a pathologist from the reference scores, categorized by the confidence of the pathologist; not certain, fairly certain and certain, from left to right. Here, we left the scores of a pathologist out of the computation of the reference scores in order to compare that pathologist to other $9$ pathologists.}
    \label{fig:data:roi:grade_distribution_deviations}
\end{figure}

We built a web-based nuclear pleomorphism platform consisting of the queries to display the ROIs, allowing the pathologists to score them through a user interface. 
Each query was followed by the questions; nuclear pleomorphism score of the tumor into one of the three categories; 1,2 or 3, and an optional field for confidence with the scoring; not certain, fairly certain, certain, as well as an additional text field for comments about the queried tumor and normal ROIs. 
An example is provided in Figure \ref{fig:data:roi:query}. % might be put in the acknowledgements.
In the figure, the ROIs were displayed on a smaller resolution. 
When an ROI is selected, the full resolution of the ROI was displayed with a total size of $2560 \times 2560$ pixels. 
We invited $10$ pathologists from several countries with varying levels of expertise in breast pathology to participate in this study. 
As a result, each tumor ROI, ${\hat{R}_{t}}$, was scored for nuclear pleomorphism $G=10$ times, denoted by the set of scores $\{ {y^1_{t}, y^2_{t}, \ldots, y^G_{t}} \}$.
For a tumor ROI, ${\hat{R}_{t}}$, in the query $Q_{t}$, we aggregated the pleomorphism scores of the pathologists by average pooling, referred to as the reference score of ${\hat{R}_{t}}$ and is denoted by $y^{ref}_t = \frac{1}{G} \sum^{G}_{g=1} y^g_{t}, t \in \{1,2, \ldots, T\}$.
% [{S^1_{t}, S^2_{t}, \ldots, S^G_{t}}] ) $. 
We present the reference scores as well as the pleomorphism scores of the individual pathologists for the $125$ queries in Figure \ref{fig:data:roi:grade_distribution}.
Additionally, we provide the distribution of the scores of the pathologists with respect to the reference scores, categorized by their confidence scores in Figure \ref{fig:data:roi:grade_distribution_deviations}.
Throughout our experiments, we train and validate the \AI on a large subset of this data set, using the reference scores as reference standard. 
The rest of this data set was used to evaluate the ROI-level performance of the \AI compared to the pathologists. 
A more detailed breakdown of the training, validation and evaluation subsets is provided in the experimental settings section.
\begin{table}
\centering
    \begin{tabular}{|l|cc|c|}
        \hline
        % \multirow{2}{*}{\textbf{Invasive carcinoma type}}
         & \multicolumn{2}{c|}{\textbf{ROI-study}} & \textbf{Slide-study} \\
        \cline{2-4}
        \textbf{Invasive carcinoma type} & Slides & ROIs & Slides \\
        \hline \hline
        No special type (Ductal)      & 26       & 83    & 91  \\
        Lobular                       & 6        & 21    & 22  \\
        Metaplastic carcinoma         & 3        & 8     & -   \\
        Invasive micropapillary       & 1        & 4     & 2   \\
        Mucinous                      & 1        & 3     & -   \\
        Tubular                       & 1        & 3     & 2   \\
        Malignant adenomyoepithelioma & 1        & 3     & -   \\
        Cribriform                    & -        & -     & 1   \\
        \hline \hline
        \textbf{Total}                         & 39       & 125   & 118 \\
        \hline
    \end{tabular}
    \caption{Distribution of the tumor types in the ROI and slide studies.}
    \label{tab:data:tumor_type}
\end{table}

\paragraph{Slide-study for the evaluation of the \AI.}
We collected an additional set of H\&E stained whole slide images from the cohort B,  $\mMS{V}=\{\mSS{V}_1, \mSS{V}_2, \ldots, \mSS{V}_Z\}$, with a total number of $Z=118$ slides.
The slides selected for this study did not share any overlap with the ROI-study, as the patients in each study were unique.
The selection criteria for the slides followed a similar approach to ROI-study, where we collected a large variety of slides with diverse tumor cell morphology. 
We used the available clinical data of the patients to approximate the distribution in the clinical practice. 
IDC was the most prevalent type of cancer with $71$ slides, whereas the second most was Invasive Lobular Carcinoma (ILC) with $14$ slides.
The full distribution of the tumor types in the slide-study is presented in Table \ref{tab:data:tumor_type}.

\begin{figure}
\centering
    \begin{minipage}{0.6\textwidth}
        \begin{minipage}{0.31\columnwidth}
            \centering
            \includegraphics[width=\columnwidth]{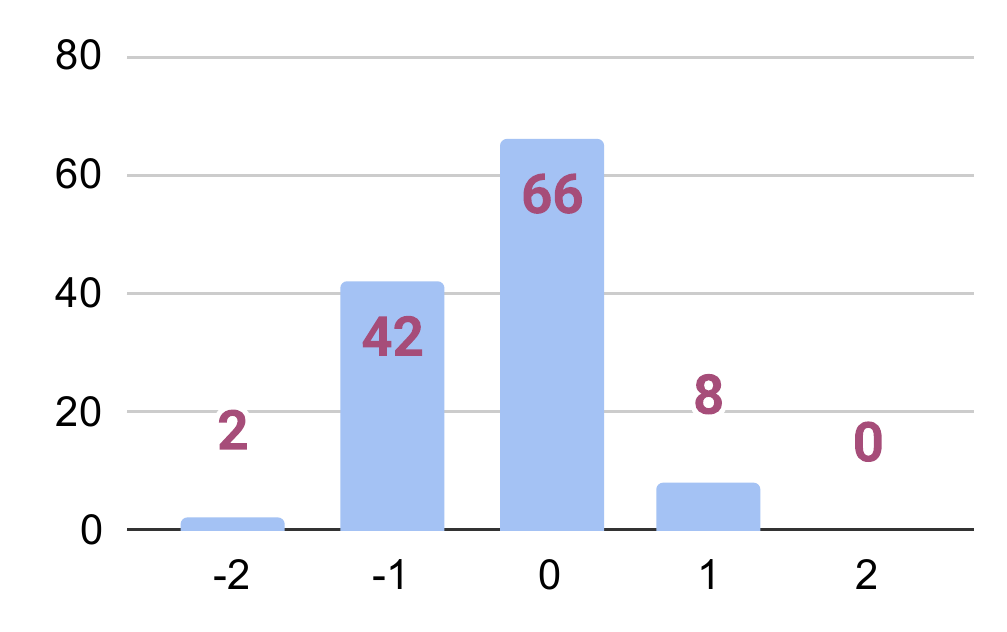}
            \subcaption{$P_{i}$ vs. $P_{ii}$}\label{fig:data:slide:pathologists_scores_difference12}
        \end{minipage}%%%
        \begin{minipage}{0.31\columnwidth}
            \centering
            \includegraphics[width=\columnwidth]{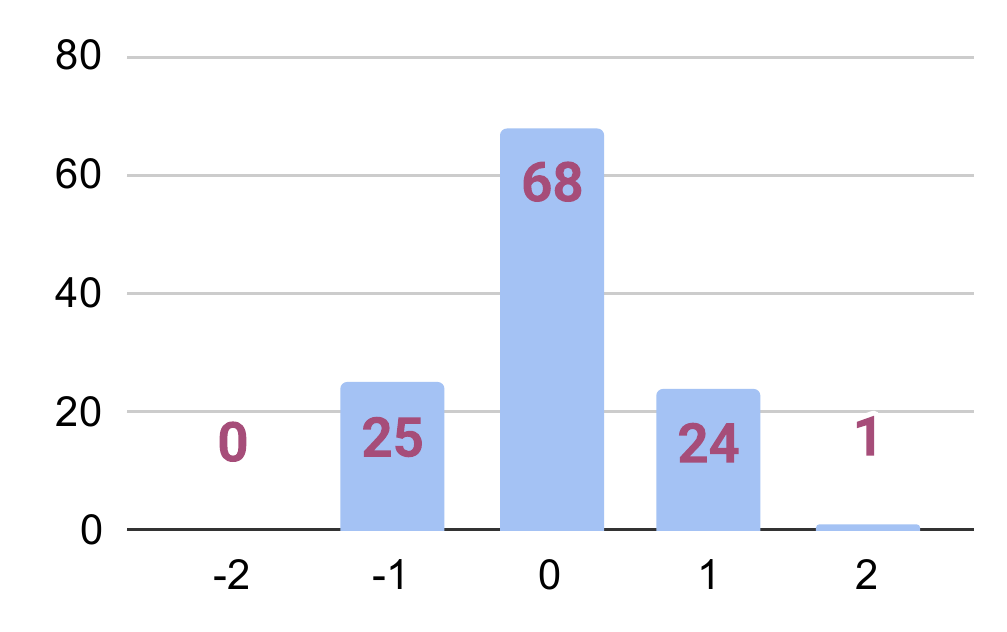}
            \subcaption{$P_{i}$ vs. $P_{iii}$}\label{fig:data:slide:pathologists_scores_difference13}
        \end{minipage}%%%
        \begin{minipage}{0.31\columnwidth}
            \centering
            \includegraphics[width=\columnwidth]{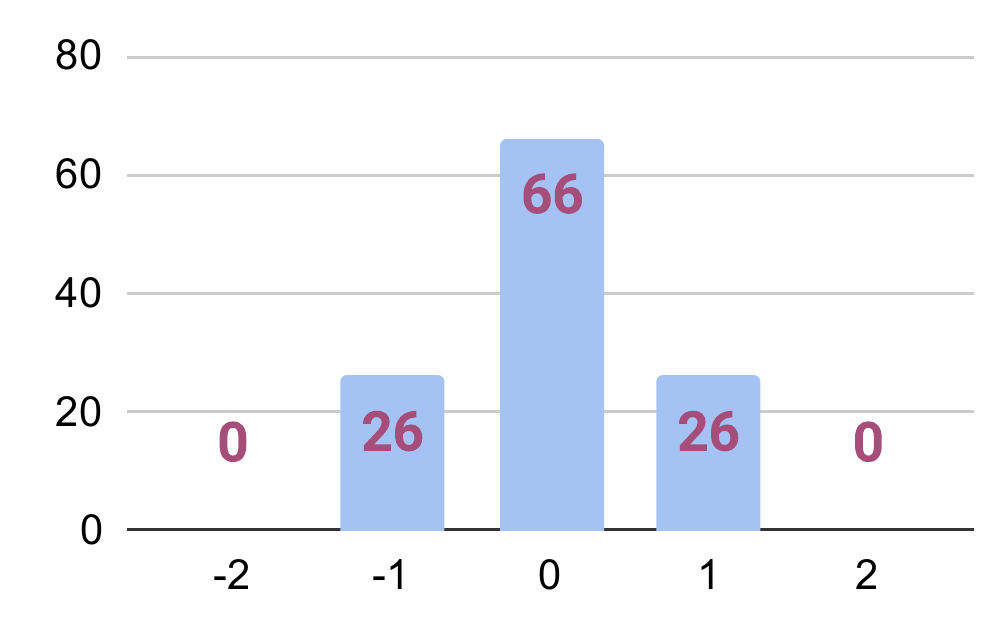}
            \subcaption{$P_{i}$ vs. $P_{iv}$}\label{fig:data:slide:pathologists_scores_difference14}
        \end{minipage}%%%
        \\
        \begin{minipage}{0.31\columnwidth}
            \centering
            \includegraphics[width=\columnwidth]{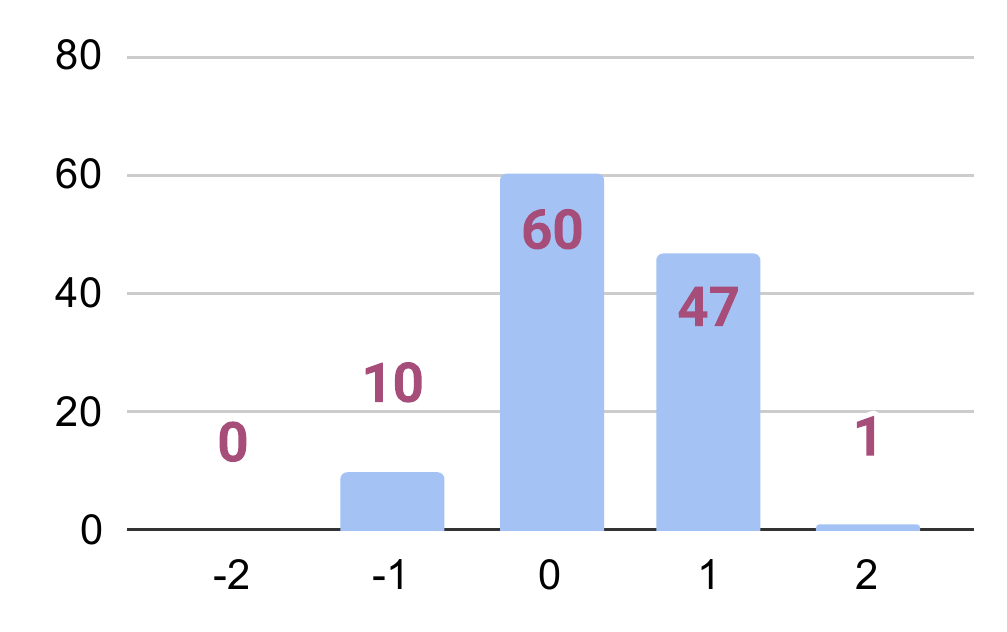}
            \subcaption{$P_{ii}$ vs. $P_{iii}$}\label{fig:data:slide:pathologists_scores_difference23}
        \end{minipage}%%%
        \begin{minipage}{0.31\columnwidth}
            \centering
            \includegraphics[width=\columnwidth]{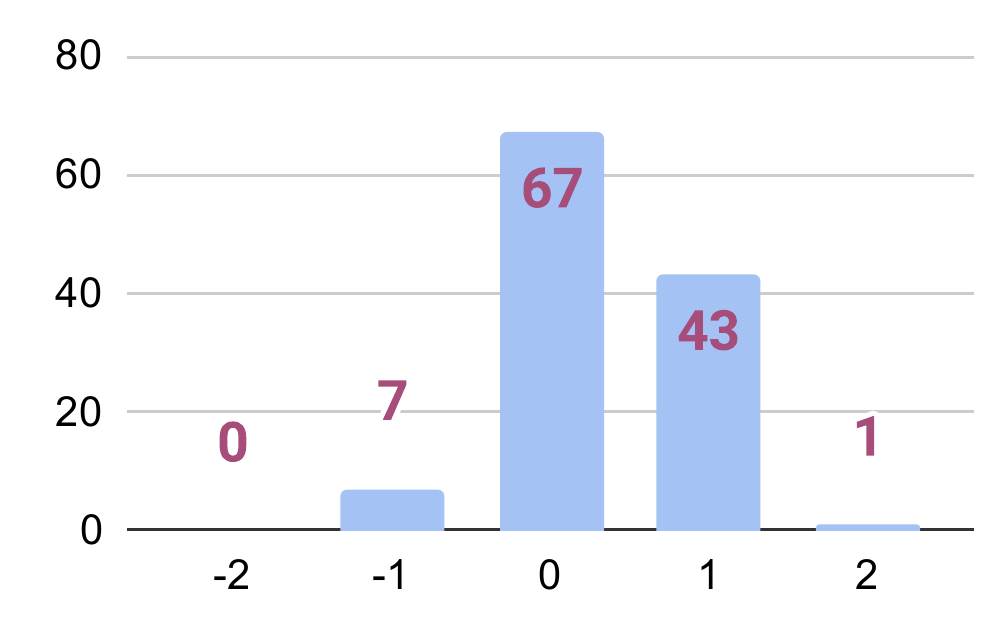}
            \subcaption{$P_{ii}$ vs. $P_{iv}$}\label{fig:data:slide:pathologists_scores_difference24}
        \end{minipage}%%%
        \begin{minipage}{0.31\columnwidth}
            \centering
            \includegraphics[width=\columnwidth]{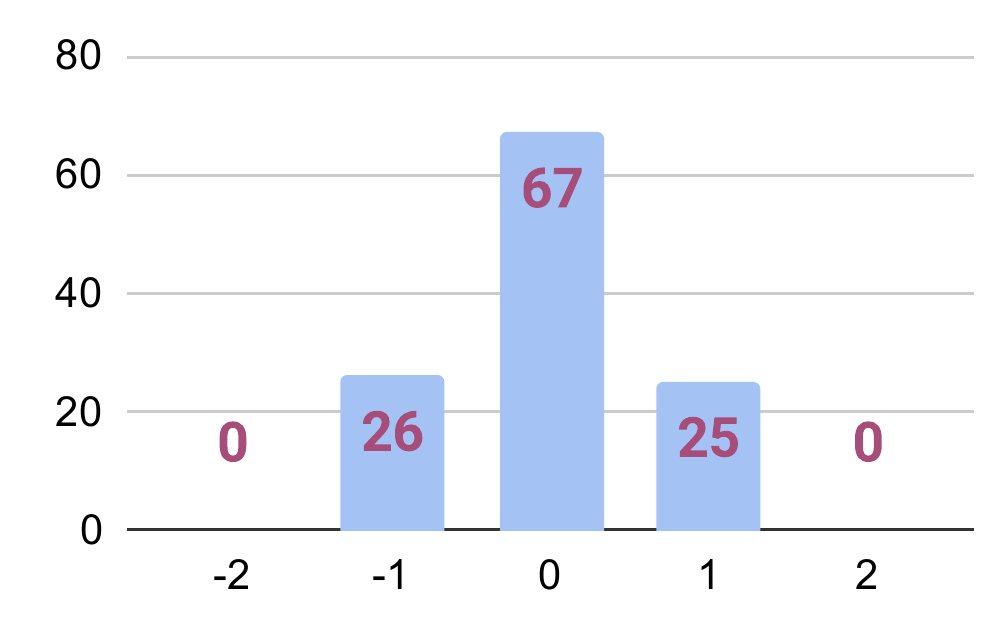}
            \subcaption{$P_{iii}$ vs. $P_{iv}$}\label{fig:data:slide:pathologists_scores_difference34}
        \end{minipage}
    \end{minipage}%%%
    \hfill\vline\hfill
    % \hspace{2pt}
    \begin{minipage}{0.35\textwidth}
    \centering
        \begin{minipage}{0.475\columnwidth}
            \centering
            \includegraphics[width=\columnwidth]{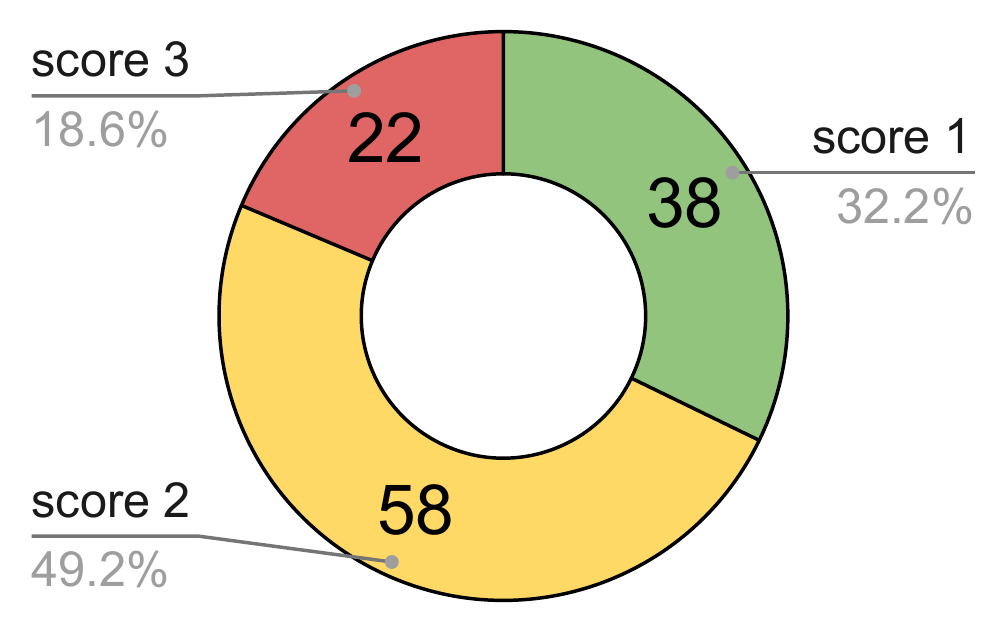}
            \subcaption{$P_i$}\label{fig:data:slide:P1_grade_distribution}
        \end{minipage}%%%
        \begin{minipage}{0.475\columnwidth}
            \centering
            \includegraphics[width=\columnwidth]{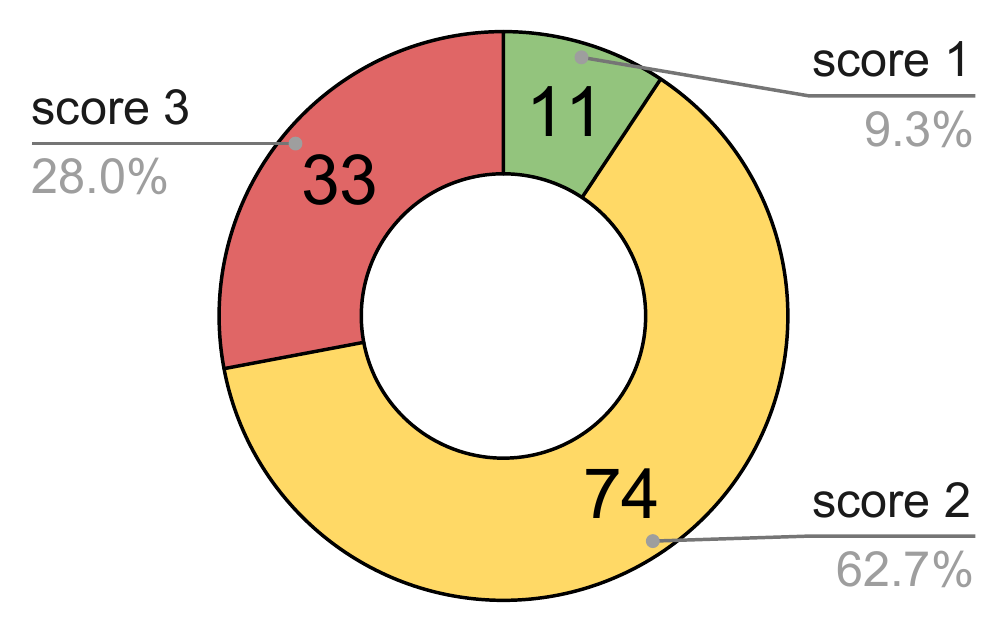}
            \subcaption{$P_{ii}$}\label{fig:data:slide:P2_grade_distribution}
        \end{minipage}
        \\
        \begin{minipage}{0.475\columnwidth}
            \centering
            \includegraphics[width=\columnwidth]{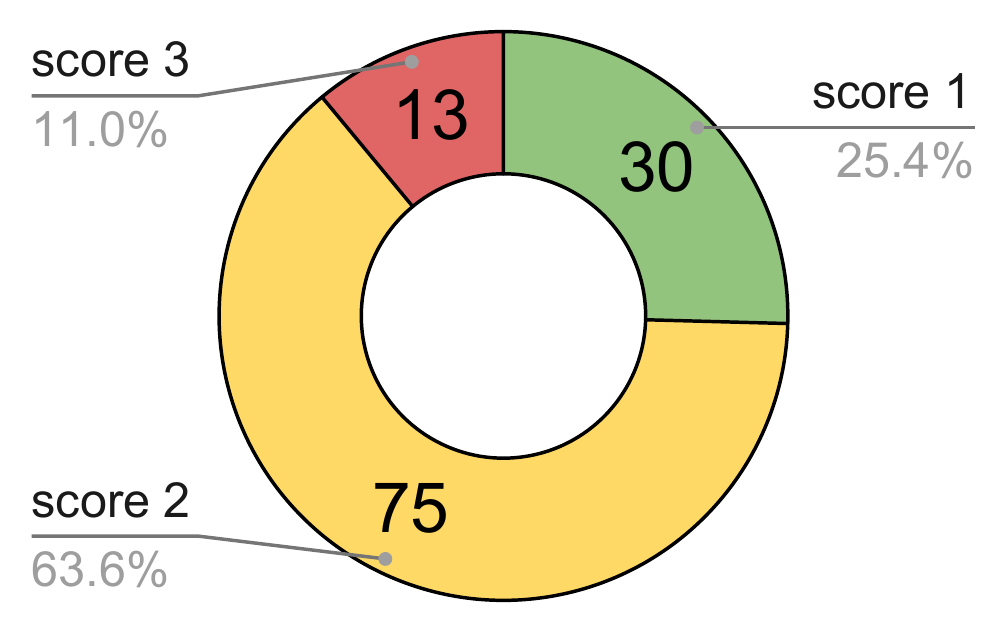}
            \subcaption{$P_{iii}$}\label{fig:data:slide:P3_grade_distribution}
        \end{minipage}%%%
        \begin{minipage}{0.475\columnwidth}
            \centering
            \includegraphics[width=\columnwidth]{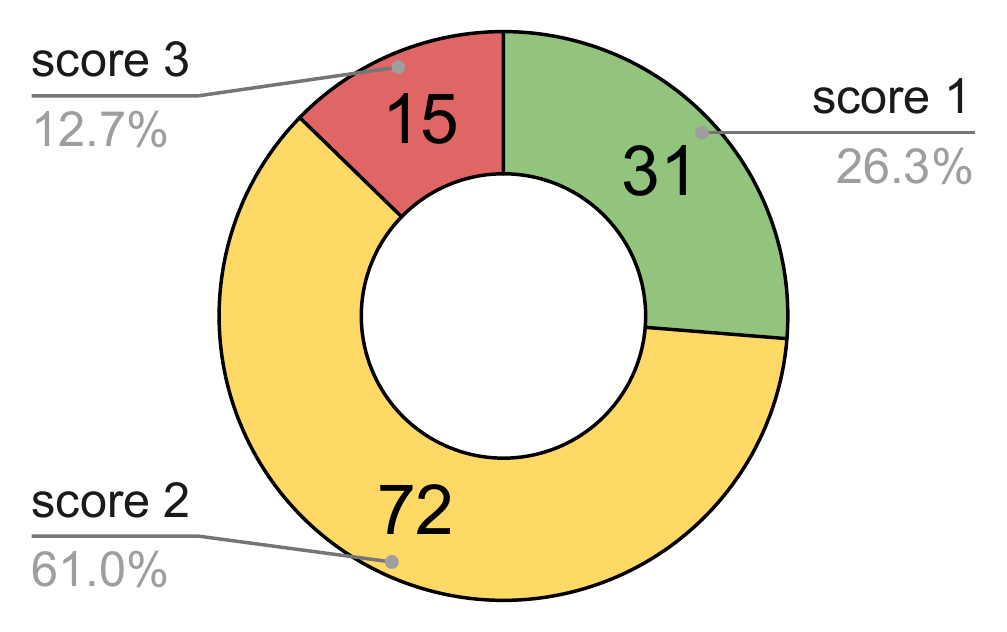}
            \subcaption{$P_{iv}$}\label{fig:data:slide:P4_grade_distribution}
        \end{minipage}
    \end{minipage}
\caption{The comparison of the pleomorphism scores of the four pathologists (a)-(f) as well as the score distribution of the pathologists (g)-(j) in the slide-study display the scoring patterns of the pathologists on the $118$ whole slide image in the slide-study. The largest differences were observed between $P_{ii}$ and other pathologists, with $P_{ii}$ assigning significantly lower number of score 1s and much higher number of score 3s compared to the other three pathologists.}
\label{fig:data:slide:pathologists_grade_stats}
\end{figure}

A web platform \footnote{www.grand-challenge.org} was built for scoring nuclear pleomorphism on the $118$ slides. 
The application could display multi-resolution whole-slide images to enable the pathologists to freely navigate the slide with zoom-in and zoom-out capabilities.
In this slide-level study, we requested from pathologists to score nuclear pleomorphism scores of whole slide images as 1,2 or 3 and provided them an optional comment field for each queried slide, as presented in Figure \ref{fig:data:slide:query}.
We invited three of the 10 pathologists who participated in the ROI-study, to also score nuclear pleomorphism in the slide-study.
Moreover, we invited an extra pathologist who did not participate in the ROI-study, bringing the total number of pathologists in this study to four.
Detailed breakdown of the score distributions of the pathologists, individually, and compared to each other is displayed in Figure \ref{fig:data:slide:pathologists_grade_stats}.
In our experiments, we used the whole slide images and the pleomorphism scores of the pathologists only to evaluate the performance of the \AI, as opposed to the ROI-study where training and validation of the \AI were also carried out.

\begin{figure}
    \centering
    \includegraphics[width=\textwidth]{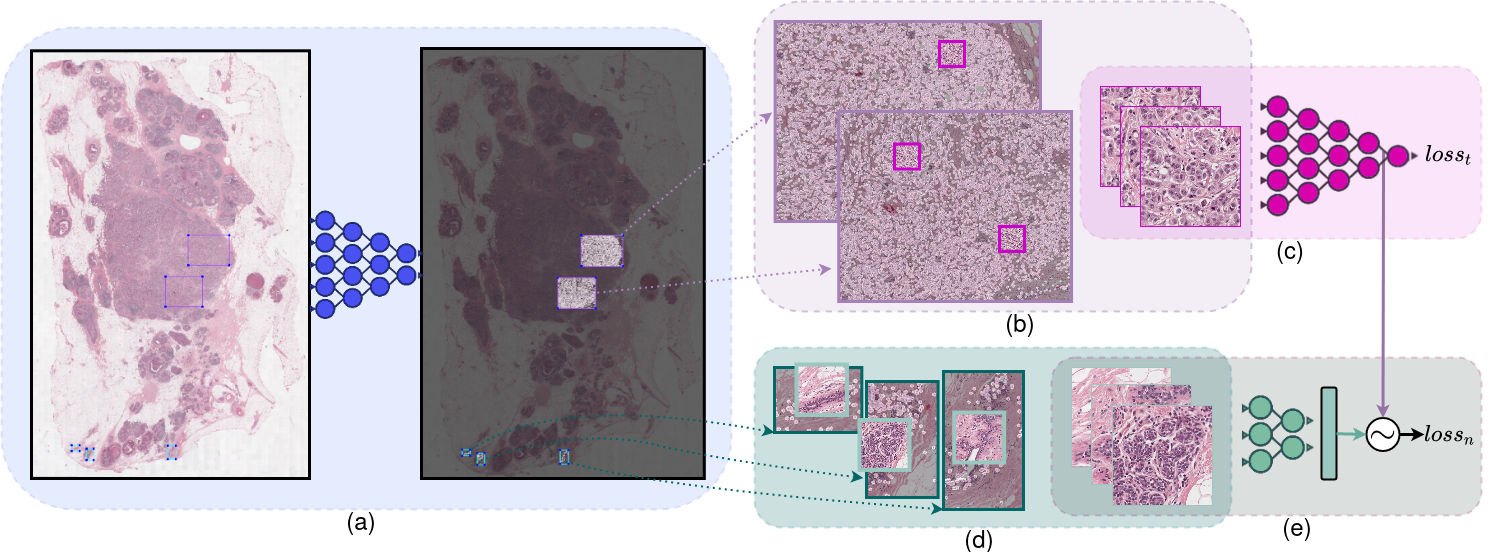}
    \caption{Proposed methodologies for scoring nuclear pleomorphism, (a)-(c) by tumor morphology as a spectrum and (a)-(e) by incorporating additional knowledge from normal epithelium. The weights of the \celldetector (a) were kept frozen during the training of the \network (c) for regressing pleomorphism spectrum as well as during the joint training using the \networkembedding (e) for the investigation of the added value of normal epithelium as baseline.}
    \label{fig:methodology:training}
\end{figure}

\paragraph{A two-stage approach for nuclear pleomorphism scoring.}
We propose a two-stage methodology to score nuclear pleomorphism on whole slide images in breast cancer, following the routine practice in which pathologists analyse the tumor in several cancer zones and compare the morphology of the tumor to the morphology of the cells in normal epithelium. 
Therefore, the first stage consists of the detection of tumor and normal cells using our \celldetector.
For the second stage, we propose to train a \network to learn nuclear pleomorphism scoring as a spectrum, considering it as a continuum instead of the traditional three-category classification.
Additionally, we propose an extension to the \network to discuss the additional benefit of incorporating knowledge from normal epithelium.
An illustration of the proposed two-stage methodology as well as the extension to the \network is presented in Figure \ref{fig:methodology:training}.

\paragraph{Tumor detection and patch sampling by tumor density.}
In the first stage of our methodology, we use the \celldetector to find the tumor and normal cells in the ROIs from the whole slide images in the ROI-study. 
This network is based on one of the state-of-the-art single-shot object detection networks, RetinaNet \cite{Lin17}, which we had trained on annotated nuclei for the detection of tumor and normal cells in our previous work.
In this work, we use this network to ensure that the patches sampled from the ROIs during the training of the pleomorphism scoring network contains high cellular activity.
It is particularly important to ensure tumor density in the sampled patches for the \network to learn to make its predictions based on tumor cells, and not on other factors in the tissue.
For inference, the \celldetector is used to process whole slide images to detect and segment the tumor regions for the \network to score nuclear pleomorphism only in diagnostically relevant regions.
Throughout the learning procedure, \celldetector was only used for inference purposes keeping the network weights frozen.

\paragraph{Nuclear pleomorphism as a continuum.}
Nuclear pleomorphism is traditionally scored into one of the three categories, 1,2 or 3, according to the grading criteria. 
Even though the clinical practice mostly follows this scoring criterion, pathologists can introduce more sensitivity by adding or subtracting half a score when the pleomorphism of the tumor is difficult to categorize. 
Motivated by this practice in clinic and our observation from the ROI-study (Figure \ref{fig:data:roi:grade_distribution}), we propose a more generalized version of the categorical classification by considering the nuclear pleomorphism as a continuous spectrum of the change in tumor morphology.
This realization transforms the three-category classification problem into a regression task on the full nuclear pleomorphism spectrum.
The granularity in the pleomorphism spectrum was made possible by the averaged collective knowledge of the $10$ pathologists.
Therefore, the second stage in our methodology is comprised of a \network based on Densenet \cite{Huan17}, which is trained on patches sampled by tumor density from the ROIs using the average (reference) scores of the pathologists as reference standard. 
A Densenet architecture consists of the so-called 'dense blocks' in which the input $x^{(l)}$ of a layer $l$ is the concatenation of the output of the previous block and the inputs of the layers preceding it, $x^{(l)}=[x, f^1(x), f^2([f^1(x) ,x]), \ldots, f^{l}([f^{l-1}([\ldots, x]) ,x])]$.
Another distinguishing feature of Densenet was the addition of the so-called 'transition layers', which help to regulate the number of channels in the network.
In comparison, a traditional CNN is only composed of consecutive layers, in which an input of a layer is the output of the previous layer, $x^{(l)}=f^{l-1}(f^{l-2}(\ldots(x)))$. 
A training iteration of our Densenet based \network is as follows.
A training patch $\hat{x}_t^i$, sampled from the tumor ROI, $\hat{R}_t$, is fed through the \network to regress its pleomorphism score $f(\hat{x}_t^i) \in [1-\epsilon, 3+\epsilon]$.
The smoothL1 loss function is computed from the predicted value $f(\hat{x}_t^i)$ with respect to the reference score $y^{ref}_t \in [1,3]$ as
\begin{align*}
loss_t &= smooth_{L1} ( f(\hat{x}_t^i) - y^{ref}_t ),  \\
smooth_{L1} (x) &= 
    \begin{cases}
        |x| &\quad\text{if}~ |x| > \alpha, \\
        \frac{1}{\alpha} x^{2} &\quad\text{if}~ |x| \leq \alpha
    \end{cases}
\end{align*}
where $\alpha$ is the hyperparameter of the loss function, corresponding to the degree of the contribution of $L_1$ and $L_2$ losses, which we set to the default value of $1$ in our experiments.
Finally, the gradient of the loss function is backpropagated through the network to improve the prediction performance in the next iteration by updating the network weights. 
This training iteration is done on a batch of multiple such patches from various tumor ROIs in the training set, $\mMI{\hat{x}} = \{\hat{x}^1, \hat{x}^2, \ldots, \hat{x}^B\}$, with $B$ denoting the batch size. 
This process of one iteration in training is visualized in Figure \ref{fig:methodology:training} through (a) to (c).
Tens of thousands of patches sampled from the tumor ROIs in the training set as well as the granularity of the reference nuclear pleomorphism scores make it possible for the network to analyse a diverse tumor morphology as a spectrum.  

\paragraph{Added value of normal epithelium as baseline.}
Following the routine clinical workflow in which normal epithelium is used as baseline for the nuclear pleomorphism scoring of tumor, we propose an extension to our \network in order to investigate the additional value that can be leveraged from the morphology of normal cells.
In this extended \network, the tumor patches, $\mMI{\hat{x}} = \{\hat{x}^1, \hat{x}^2, \ldots, \hat{x}^P\}$, are used as input to the Densenet architecture, as before. 
The normal patches, $\mMI{\bar{x}} = \{\bar{x}^1, \bar{x}^2, \ldots, \bar{x}^P\}$, are fed through a simpler \networkembedding, denoted by $g$, as normal cells have little to no variability in morphology compared to the large plethora of the tumor morphology.
The architectural details of the \networkembedding are outlined in Table \ref{tab:method:normal:network}.
In this extension, the feature representation from a tumor patch is compared to the feature embedding from a normal patch through a cosine similarity loss function
\begin{align*}
\centering
loss_n &= max \left( 0, \left[ \phi(y^{ref}_t) ~cos \left( f^p(\hat{x}_t^i), g(\bar{x}_n^j) \right) \right]  \right), \\
\phi(y^{ref}_t) &= 0.5 y^{ref}_t - 1
\end{align*}
where $y^{ref}_t \in [1,3]$ is the reference score of the tumor patch $\hat{x}_t$, $f^p$ denotes the penultimate layer of the \network from which the feature representation of the tumor patch is obtained.
This loss function ensures that the feature representations of the tumor patches with reference scores around $1$ will resemble the most to the normal patches, and the patches with reference scores closer to $3$ will resemble the least.
More specifically, the feature representation of a tumor patch with size $1024$ from the penultimate layer of the \network is compared to the output of a normal patch with the same size from the \networkembedding.
The cosine similarity function is parameterized by a function of the reference score of the tumor patch to control the penalization with respect to the similarity versus the normal patch.
During learning, both the \network and the \networkembedding are trained jointly, utilizing both loss functions $loss_t$ and $loss_n$.
During inference, only the \network is used to predict a pleomorphism score for an input image patch.
Similar to the first approach, the \celldetector is employed to ensure high cellular density in sampled tumor and normal patches.
The rest of the training procedure such as the selection of hyperparameters were kept the same as before.
This extended \network is visualized in Figure \ref{fig:methodology:training} through (a) to (e).
\begin{table}
    \centering
    \begin{tabular}{|c|l|}
    \hline
    \textbf{Layer} & \textbf{Feature embedding network} \\
    \hline \hline
    1 & conv(7,3,0)-32 + maxpool(2,2) + ReLU + BN \\
    2 & conv(5,2,0)-64 + maxpool(2,2) + ReLU + BN \\
    3 & conv(3,3,0)-64 + maxpool(2,2) + ReLU + BN \\
    4 & conv(3,3,0)-128 + maxpool(2,2) + ReLU + BN \\
    5 & fc(512,256) + ReLU \\  
    6 & fc(256,1024) + ReLU \\  
    \hline 
    \end{tabular}
    \caption{The \network was extended by the \networkembedding to investigate the added value that can be learned from normal epithelium. conv($k,s,p$)-$c_o$ is a convolutional layer with kernel size $k$, stride of $s$, padding of $p$, with $c_o$ number of output channels, maxpool($k,k$) is the max pooling operation with kernel size of $k$, ReLU corresponds to rectified linear unit, and BN refers to batch normalization. fc($c_i,c_o$) is a fully connected layer with $c_i$ number of input and $c_o$ number of output channels. }
    \label{tab:method:normal:network}
\end{table}

\paragraph{Scoring nuclear pleomorphism on ROIs and whole slide images.} 
As the input to the \AI was of size $512 \times 512$ pixels, the ROIs were too large to be processed directly.
Therefore, a single pleomorphism score for an ROI was obtained by the procedure below.
The ROI was processed into fixed-size overlapping tiles. 
In our experiments, we utilized an overlap value of $448$ pixels, horizontally and vertically, for the tile size of $512 \times 512$ pixels.
The \AI predicted a continuous pleomorphism score for each tile. 
As a result of the overlapping tiles, multiple predictions were made in small blocks of size $64 \times 64$ and for each block in the ROI, predictions were aggregated by average pooling to assign its pleomorphism score.
Subsequently, we fine-tuned the predictions by only assigning them to the tumor cells using the \celldetector.
Finally, the pleomorphism score of the ROI was determined by the average pleomorphism score in the tumor.
Whole slide-level inference followed a similar procedure in which the pleomorphism score of a slide was determined by averaging the pleomorphism scores over its tumor regions.
Additionally, a preprocessing step was applied to locate and process only the tumor regions in the slide, using the output of the \celldetector to avoid fatty tissue, stroma, normal epithelium and several other structures that are not taken into account in breast cancer grading.
Therewith, we ensured that the \network scored pleomorphism only on tumor regions.

\paragraph{Experimental Setting.}
The majority of the data set collected in the ROI-study was mainly used for the training and the validation of the \AI, and only a small subset was used for evaluation.
On the other hand, the entire set of the whole slide images in the slide-study were only used for evaluation.
The selection of the slides for the training and the validation of the \AI was carried out with the goal of exposing it to as much variety of tumor morphology as possible.
As a result, we selected the 52 ROIs from 16 slides and the 28 ROIs from 10 slides in the ROI-study.
The remaining 45 ROIs from the 13 slides were used to evaluate the performance of the \AI. 
The performance evaluation in the ROI-study could be regarded as the result of a controlled experiment in which each test ROI was homogeneous in nuclear pleomorphism severity.
The 118 slides from the slide-study were only used for slide-level performance evaluation of the \AI compared to the pathologists.
This setting represented the real-world clinical practice where the \AI scored nuclear pleomorphism on whole slide images.

We trained the Densenet based \network from scratch using the Adam optimizer with a learning rate of $1e-4$.
Each training patch was $512 \times 512$ pixels, extracted from the tumor ROIs in the training set of the ROI-study slides at $0.5$ spacing.
The batch size used during an iteration of the training was $12$.
Each epoch contained 200 training and 500 validation iterations.
Training of the \network continued until the loss value in the validation set no longer improved. 
% Data augmentation played a great role in the performance of the \AI.
The training and validation patches were augmented with spatial operations such as horizontal and vertical flips and $90^{\circ}, 180^{\circ}, 270^{\circ}$ rotations.
In order to make the \AI color and stain invariant, we applied heavy color and stain augmentations on the training patches \cite{Tell19}.
The training patches were also augmented with blurring techniques to make the network more robust against out-of-focus artifacts, commonly seen in whole slide images.
A Gaussian map was applied on the output of the \celldetector to create a density map of the tumor cells which was used to sample patches from areas with high cellular density during training.
A batch had a balanced pleomorphism distribution among its patches, which was made possible by an equal distribution of quantized reference scores.

The performance of the \network was compared to the performance of the extended \network to investigate the added value of normal epithelium as a baseline, using several regression metrics, such as mean absolute error ($MAE$), mean squared
error ($MSE$) and median absolute error ($MdAE$), explained variance score ($EV$) and $R^2$ score.
Throughout the other quantitative evaluations, the predictions of the \AI were quantized into several categories, most notably into the traditional three-category setting to compare the performance against the pathologists.
To illustrate, quantization of the range $[1,3]$ into three categories was to partition the score spectrum into three equal sized categories; a prediction within the range of $[1,1.66)$ corresponded to score 1, $[1.66, 2.33]$ corresponded to score 2, and $(2.33, 3]$ corresponded to score 3. 
We evaluated the performance of the \AI versus the pathologists as well as the performance of each pathologist versus one another, by Cohen's quadratic kappa measure \cite{McHu12}
\[
\kappa = \frac{p_o - p_e}{1 - p_e}
\]
where $p_o$ denotes the relative observed agreement between two participants and $p_e$ corresponds to the probability of chance agreement. 

\printbibliography

\section*{Disclosures}
Caner Mercan is supported by the NWO-Perspectief program EDL (Efficient Deep Learning), co-financed by the Dutch Organisation for Scientific Research and 35 Dutch companies.

\noindent Jeroen van der Laak is a member of the advisory boards of Philips, The Netherlands and ContextVision, Sweden, and received research funding from Philips, The Netherlands, ContextVision, Sweden, and Sectra, Sweden in the last five years.

\noindent Roberto Salgado reports non-financial support from Merck and Bristol Myers Squibb; research support from Merck, Puma Biotechnology, and Roche; and advisory board fees for Bristol Myers Squibb; and personal fees from Roche for an advisory board related to a trial-research project.

\end{document}